%% file: main.tex
\documentclass[twocolumn]{aastex62}
\usepackage{xcolor}
\newcommand{\package}[1]{\textsl{#1}}

\input{shorthand.tex}

\shorttitle{NSC DR2}
\shorttitle{Nidever et al.}

\begin{document}

\title{Second Data Release of the All-sky NOIRLab Source Catalog}
\input{authors.tex}

\begin{abstract}
We announce the second data release (DR2) of the NOIRLab Source Catalog (NSC), using 412,116 public images from CTIO-4m+DECam, KPNO-4m+Mosaic3 and Bok-2.3m+90Prime.  NSC DR2 contains over 3.9 billion unique objects, 68 billion individual source measurements, covers $\approx$35,000 square degrees of the sky, has depths of $\approx$23rd magnitude in most broadband filters with $\approx$1--2\% photometric precision, and astrometric accuracy of $\approx$7 mas. Approximately 1.9 billion objects within $\approx$30,000 square degrees of sky have photometry in three or more bands.
There are several improvements over NSC DR1.  DR2 includes 156,662 (61\%) more exposures extending over 2 more years than in DR1.  The southern photometric zeropoints in $griz$ are more accurate by using the Skymapper DR1 and ATLAS-Ref2 catalogs, and improved extinction corrections were used for high-extinction regions.  In addition, the astrometric accuracy is improved by taking advantage of Gaia DR2 proper motions when calibrating the WCS of individual images.  This improves the NSC proper motions to $\sim$2.5 mas/yr (precision) and $\sim$0.2 mas/yr (accuracy).  The combination of sources into unique objects is performed using a DBSCAN algorithm and mean parameters per object (such as mean magnitudes, proper motion, etc.) are calculated more robustly with outlier rejection.  Finally, eight multi-band photometric variability indices are calculated for each object and variable objects are flagged (23 million objects).  NSC DR2 will be useful for exploring solar system objects, stellar streams, dwarf satellite galaxies, QSOs, variable stars, high-proper motion stars, and transients. The NSC DR2 catalog is publicly available via the NOIRLab's Astro Data Lab science platform.
\end{abstract}

\keywords{surveys - catalogs}




\section{Introduction} \label{sec:intro}

In the last twenty years large and systematic digital imaging surveys of the sky have revolutionized astronomical exploration. Beginning with the Sloan Digital Sky Survey  \citep[SDSS;][]{York2000}, ground-based surveys like the PS1 \citep[Pan-STARRS1][]{Chambers2016}, Dark Energy Survey
 \citep[DES;][]{Abbott2017}, Legacy Surveys  \citep[LS;][]{Dey2019}, the DECam Plane Survey \citep[DECaPS;][]{Schlafly2018}, Zwicky Transient Factory \cite[ZTF;][]{Bellm2015}, and others have mapped the sky at multiple bands, epochs and cadences.
 The wealth of data from these large surveys has enabled a wide variety of discoveries and expanded our ability to explore the universe with large statistical samples.
The surveys have yielded, for example, dozens of new Milky Way satellite dwarf galaxies \citep[by DES; e.g.,][]{Bechtol2015,Drlica-Wagner2015}, to systematic searches of variable stars in the Galactic halo \citep[by PS1;][]{Sesar2017b}, and troves of supernovae in distant galaxies \citep[e.g.,][]{Perley2020}, and much more.
 In the near future, the Legacy Survey of Space and Time \citep[LSST;][]{Ivezic2008} with the Rubin Observatory will further revolutionize astronomy by mapping the southern skies every three nights for ten years. 

\begin{figure*}[!ht]
\begin{center}
\includegraphics[width=1.0\hsize,angle=0]{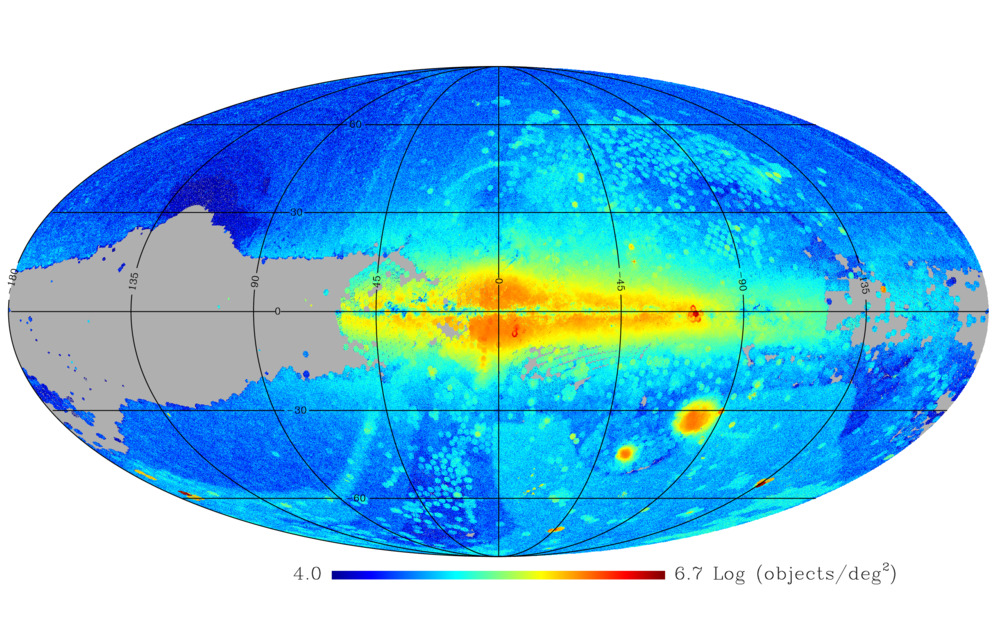}
\end{center}
\caption{Density of the 3.9 billion NSC objects on the sky in Galactic coordinates.  The higher densities from the Galactic midplane and Bulge as well as the LMC and SMC are readily apparent.  The density is a combination of the true density of objects as well as the particular exposure times of the various observing programs.}
\label{fig_bigmap}
\end{figure*}

A great resource that is often overlooked is the large wealth of public imaging data that exist in national observatory data archives.  These data are inhomogeneous, including both large systematic surveys and smaller PI-driven programs. Hence, significant effort is required in order to make the entire dataset useful to the community as a combined ``survey'', with uniform reductions and calibrations suitable for astronomical exploration.  Similar efforts been undertaken with other facilities, resulting in, e.g., the Chandra Source Catalog \citep{Evans2010} and Hubble Source Catalog \citep{Whitmore2016}. However, the variable observing conditions and large variety of telescope and instrument combinations have made this effort more formidable for ground-based optical and NIR imaging archival data.  The NOIRLab Source Catalog (NSC)\footnote{Formerly known as the NOAO Source Catalog.} is an effort to create such a uniformly processed dataset using the images in the NOIRLab Astro Data Archive\footnote{\url{https://astroarchive.noao.edu/}}.  The first data release \citep[NSC DR1;][]{Nidever2018} consisted of 2.9 billion object with 34 billion individual measurements from over 195,000 images.

Here we present the second public data release of the NSC (NSC DR2).  
It catalogs 3.9 billion unique sources, representing the largest single astronomical source catalog to date.  The 68 billion individual measurements from 412,116 images more than doubles the total data volume from NSC DR1.  Besides more data, NSC DR2 includes some important processing updates.  We use recently released wide-area catalogs (ATLAS-Refcat2, \citealt{Tonry2018}; and Skymapper DR1, \citealt{Wolf2018}) to improve our photometric calibration in the south, and more accurate extinction estimates \citep[e.g., RJCE method][]{Majewski2011} for the smaller number of model magnitudes that we still employ for zero point estimates.  The Gaia DR2 \citep{Gaia2016,GaiaDR2} astrometry and proper motion corrections are used to obtain improved astrometric calibration of the images which, in turn, produce more accurate NSC proper motion measurements.  A more sophisticated algorithm is used to group individual measurements into ``objects'' using DBSCAN clustering.  In addition, eight photometric variability metrics are computed for each object and 10$\sigma$ outliers are automatically flagged.  These enhancements improve the precision of the data, reduce systematics, and add more valuable information that will make it easier for users to exploit the data for a variety of scientific goals.

The paper is laid out in the following manner. The imaging dataset is described in Section \ref{sec:data} while a description of the data reduction and processing steps is given in Section \ref{sec:phot}.   A brief discussion of caveats are presented in Section \ref{sec:caveats}.  The overall catalog and the achieved performance and reliability are discussed in Section \ref{sec:performance}.  A number of science use cases of NSC DR2 are presented in Section \ref{sec:science}.  Finally, Section \ref{sec:summary} gives a brief summary.

\begin{center}
\begin{figure}[t]
\includegraphics[width=1.0\hsize,angle=0]{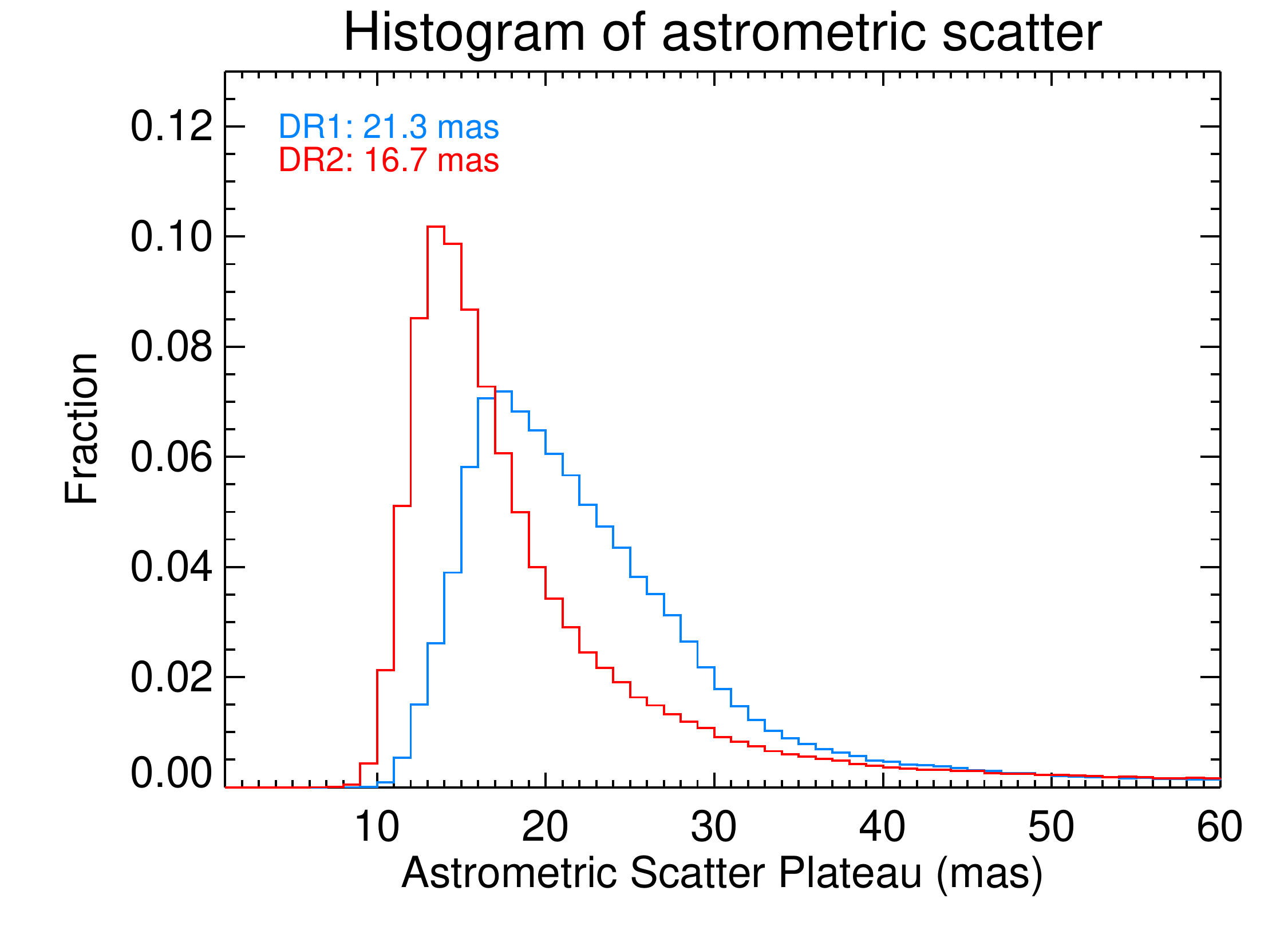}  
\caption{Histogram of rms scatter around the astrometric fit per CCD (averaged across the exposure) for DR1 and DR2.  The use of Gaia DR2 astrometry, including proper motion corrections, reduced the median scatter from 21.3 mas in the NSC DR1 to 16.7 mas in the NSC DR2. The astrometric scatter is now more tightly peaked around 14 mas.}
\label{fig_astrms}
\end{figure}
\end{center}

\section{Dataset}
\label{sec:data}

All sources in NSC DR2 are measured from public images drawn from the NOIRLab Astro Data Archive\footnote{\url{https://astroarchive.noao.edu/}}.  The majority of the images used in NSC DR2 are CTIO-4m Blanco + DECam (340,952 exposures). In addition, there are 41,561 exposures from KPNO-4m Mayall + Mosaic3 (the majority from the Mayall $z$-band Legacy Survey; MzLS; \citealt{Dey2016}) and 29,603 exposures from the Steward Observatory Bok-2.3m + 90Prime (from the Beijing-Arizona Sky Survey; BASS; \citealt{Zou2017,Zou2018,Zou2019}). A large fraction of the images are data obtained by the Dark Energy Survey \citep{Abbott2017} and the Legacy Surveys imaging projects \citep{Dey2019}.

\section{Reduction and Photometry} \label{sec:phot}

The reduction and analysis tools used are essentially the same as those used for NSC DR1 \citep[see][for details]{Nidever2018}.  We provide a brief summary here and describe the few changes.  The NSC data use images that are processed by the NOAO Community Pipelines for instrumental calibration (\citealt{Valdes2014}; Valdes et al., in preparation)\footnote{\url{https://www.noao.edu/noao/staff/fvaldes/CPDocPrelim}}.  Source Extractor\footnote{\url{https://www.astromatic.net/software/sextractor}}
\citep{Bertin1996} is used to perform source detection, aperture photometry, and morphological parameter estimation from the images.  Finally, custom software\footnote{\url{https://github.com/noaodatalab/noaosourcecatalog}} (written in Python and IDL) is used to perform photometric and astrometric calibration, to spatially cluster sources measured on different images into unique objects, and to measure their mean object properties.

The NSC processing is split into three main steps: (1) measurement, (2) calibration, and (3) combination.  These steps are described in more detail below.

\subsection{Measurement} \label{subsec:measure}

The measurement step includes detection of objects in the images, the measurement of position and aperture photometry, and the measurement of morphological parameters.  We use Source Extractor (SExtractor) with the same setup as described in \citet{Nidever2018}.  For exposures taken (and publicly available) prior to UT 2017 October 11 (the NSC DR1 cutoff date), we used the SExtractor files previously used for NSC DR1 files.  We ran SExtractor anew on exposures taken (and publicly available) after UT 2019 October 17.  
SExtractor measurement catalogs for 482,630 exposures were considered for inclusion in NSC DR2.  This was later trimmed down to 412,116 after the application of quality cuts (see section \ref{subsec:combine}).

\subsection{Calibration} \label{subsec:calibrate}

The second major NSC processing step is the astrometric and photometric calibration.  The methods are nearly identical to those used in NSC DR1 \citep[see][for details]{Nidever2018}, with two major improvements: (1) we use Gaia DR2 proper motions in the astrometric calibration, and (2) Skymapper DR1 \citep{Wolf2018} and ATLAS-Refcat2 \citep{Tonry2018} to derive photometric zeropoints for southern data (i.e., where PS1 data are not available).

\subsubsection{Astrometry}
\label{subsubsec:astrometry}

The astrometric calibration, as in NSC DR1, is performed on an exposure catalog and linear correction terms are derived using a reference catalog.  In NSC DR2, Gaia DR2 was used for the reference and the coordinates of the reference stars were precessed to the epoch of the observation using the Gaia DR2 coordinates (J2015.5) and proper motions.  Robust standard deviations of the residuals of the astrometric fit are calculated for each exposure.  The median rms of the astrometric residuals decreased from 21 mas in NSC DR1 to is 17 mas in NSC DR2, and the distributions have become much more sharply peaked (see Fig.\ \ref{fig_astrms}).  The biggest improvement is in the proper motions,  which are explained further in section \ref{sec:performance}. The rms of the {\em average} coordinates for bright stars when compared to Gaia DR2 is 7--8 mas.  

\begin{center}
\begin{figure*}[!ht]
$\begin{array}{cc}
\includegraphics[trim={0cm 4.9cm 2cm 1cm},clip,width=0.50\hsize,angle=0]{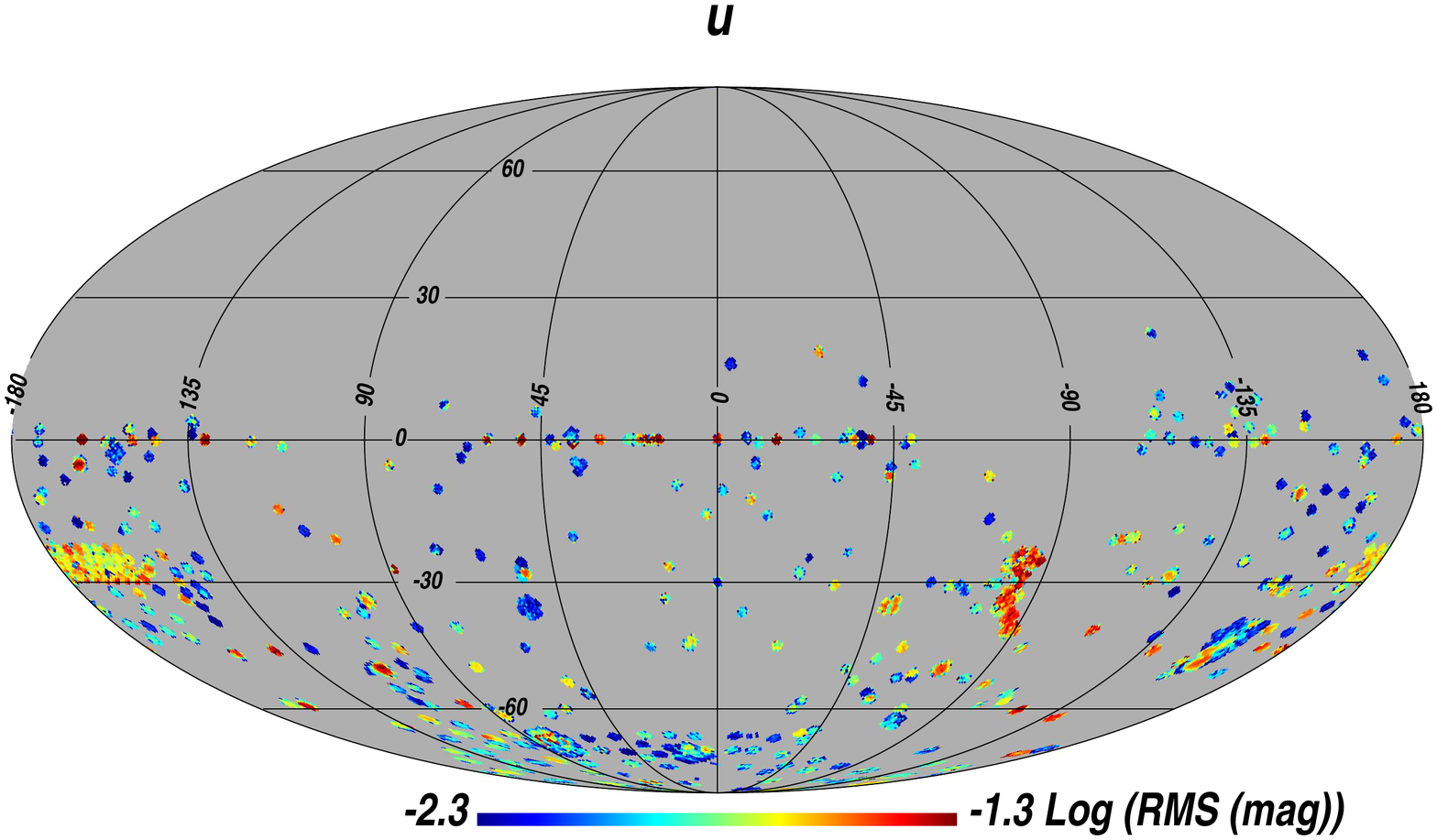}
\includegraphics[trim={0cm 4.9cm 2cm 1cm},clip,width=0.50\hsize,angle=0]{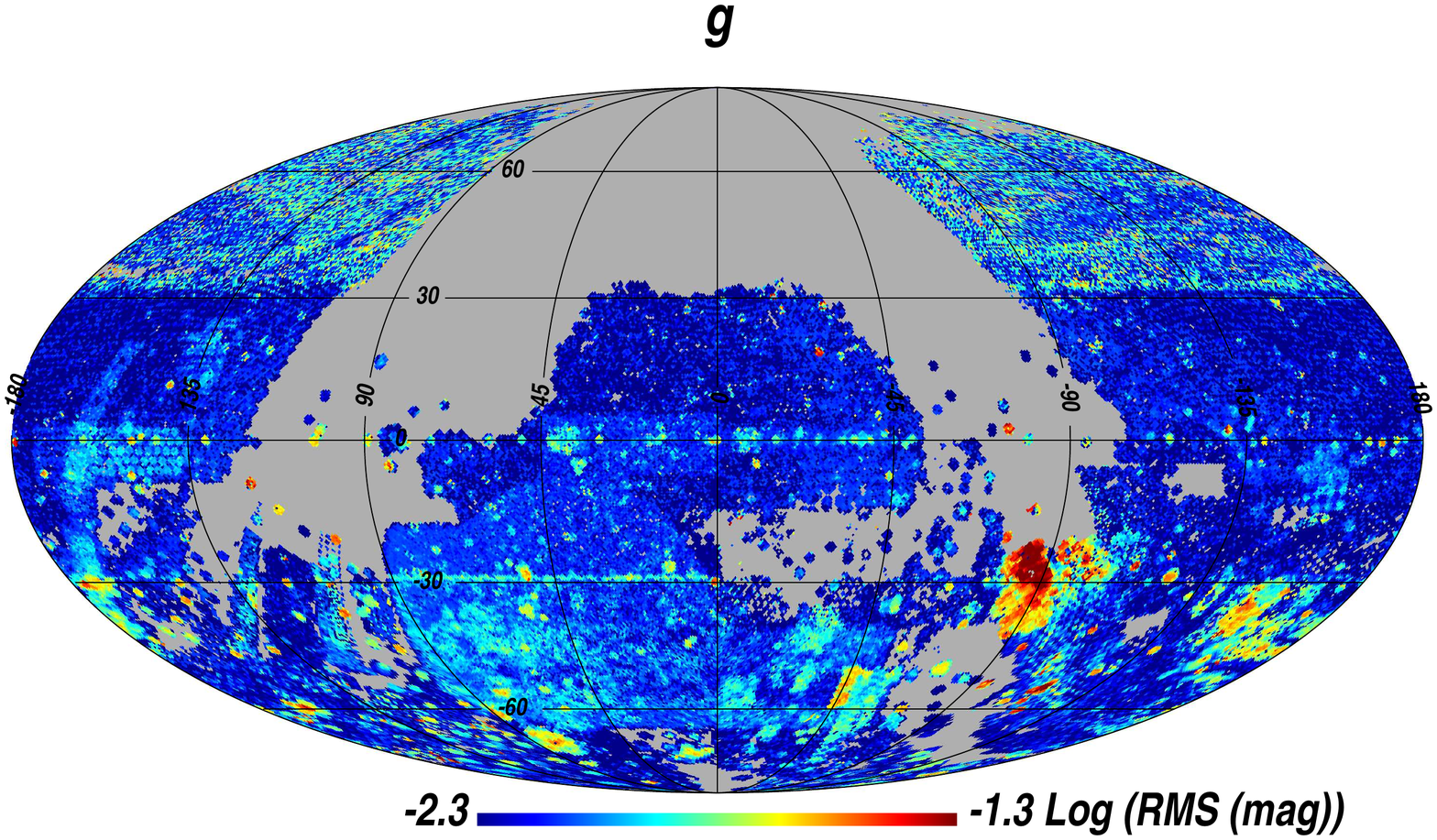} \\
\includegraphics[trim={0cm 4.9cm 2cm 1cm},clip,width=0.50\hsize,angle=0]{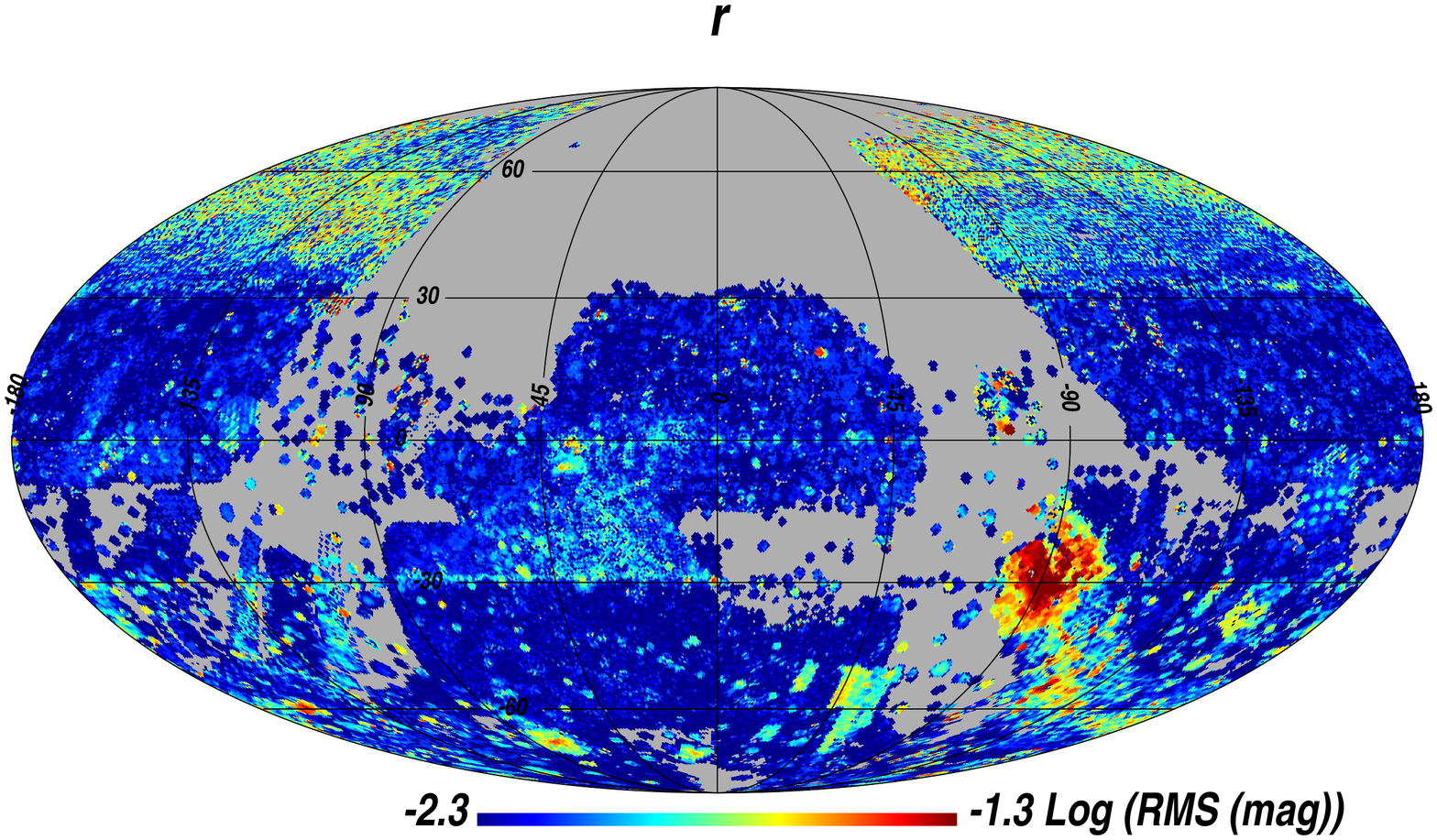}
\includegraphics[trim={0cm 4.9cm 2cm 1cm},clip,width=0.50\hsize,angle=0]{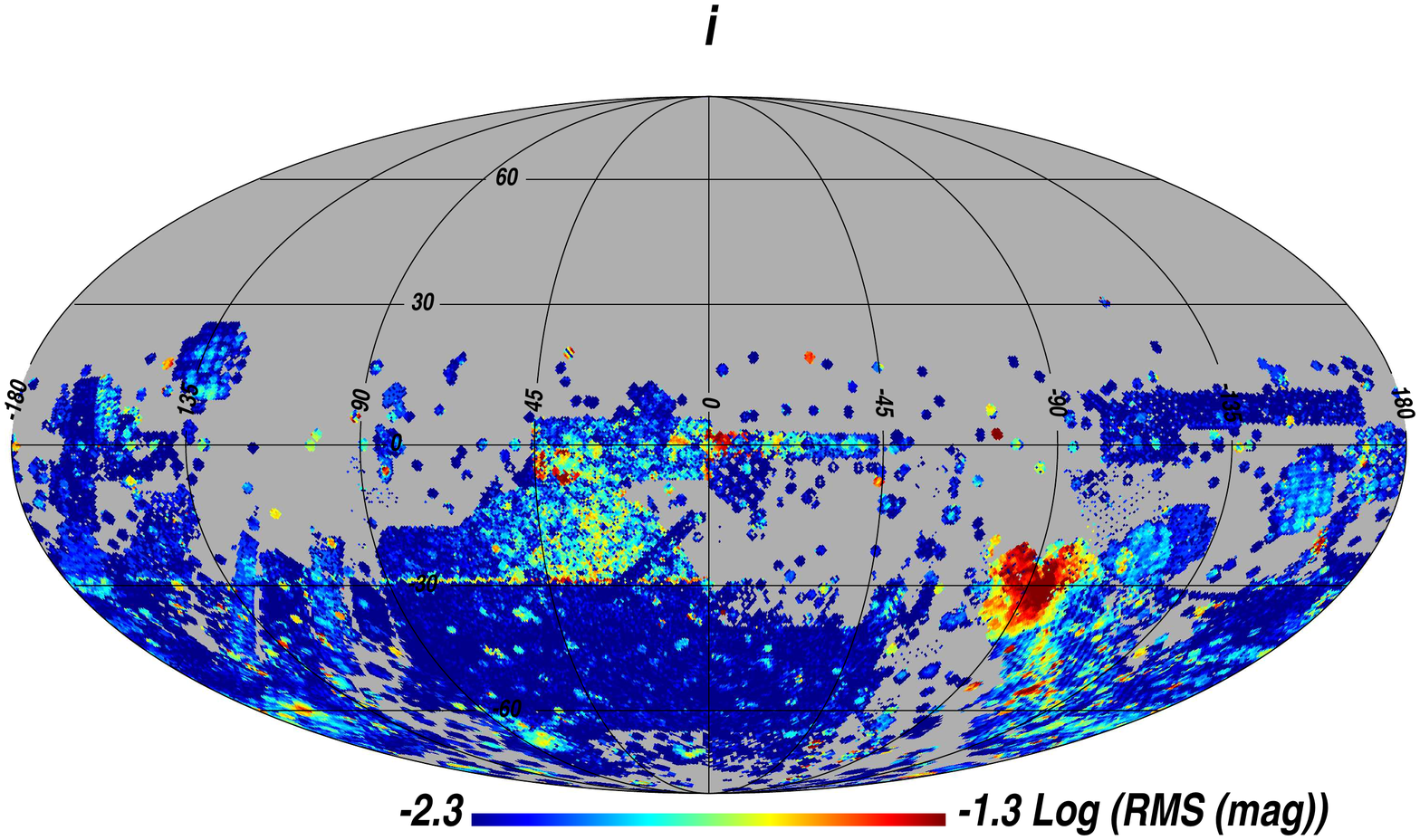} \\
\includegraphics[trim={0cm 4.9cm 2cm 1cm},clip,width=0.50\hsize,angle=0]{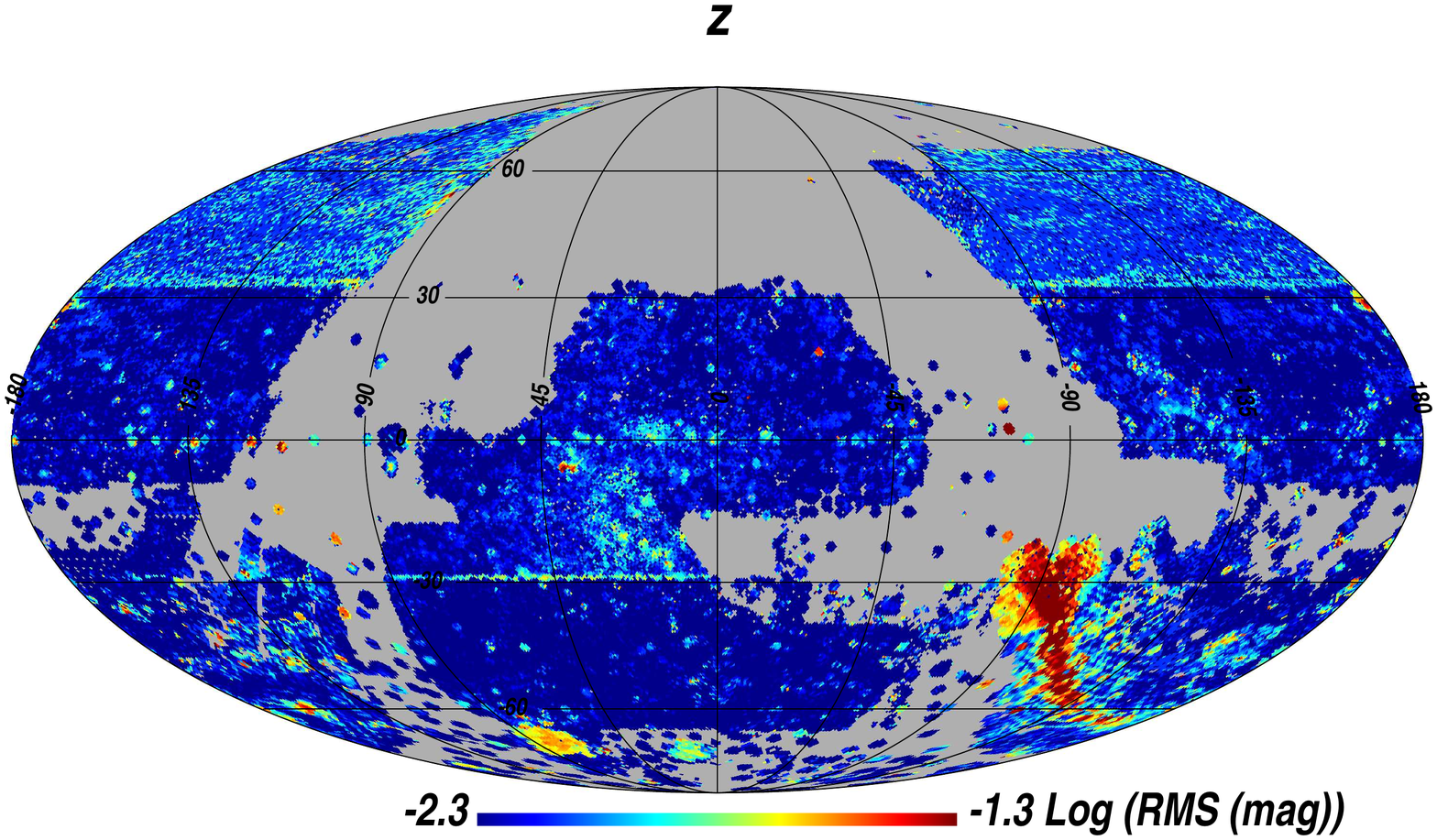}
\includegraphics[trim={0cm 4.9cm 2cm 1cm},clip,width=0.50\hsize,angle=0]{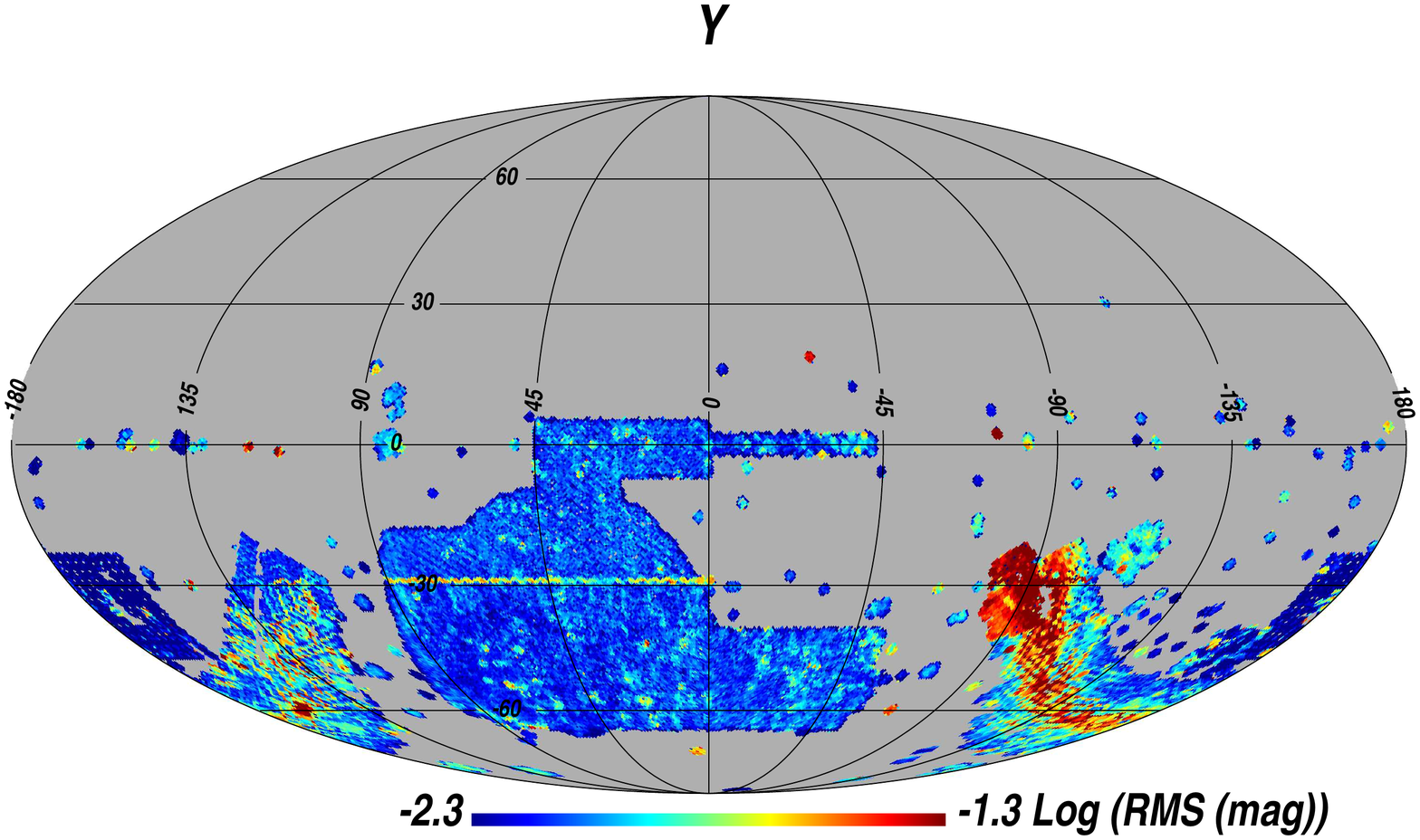} \\
\includegraphics[trim={0cm 4.9cm 2cm 1cm},clip,width=0.50\hsize,angle=0]{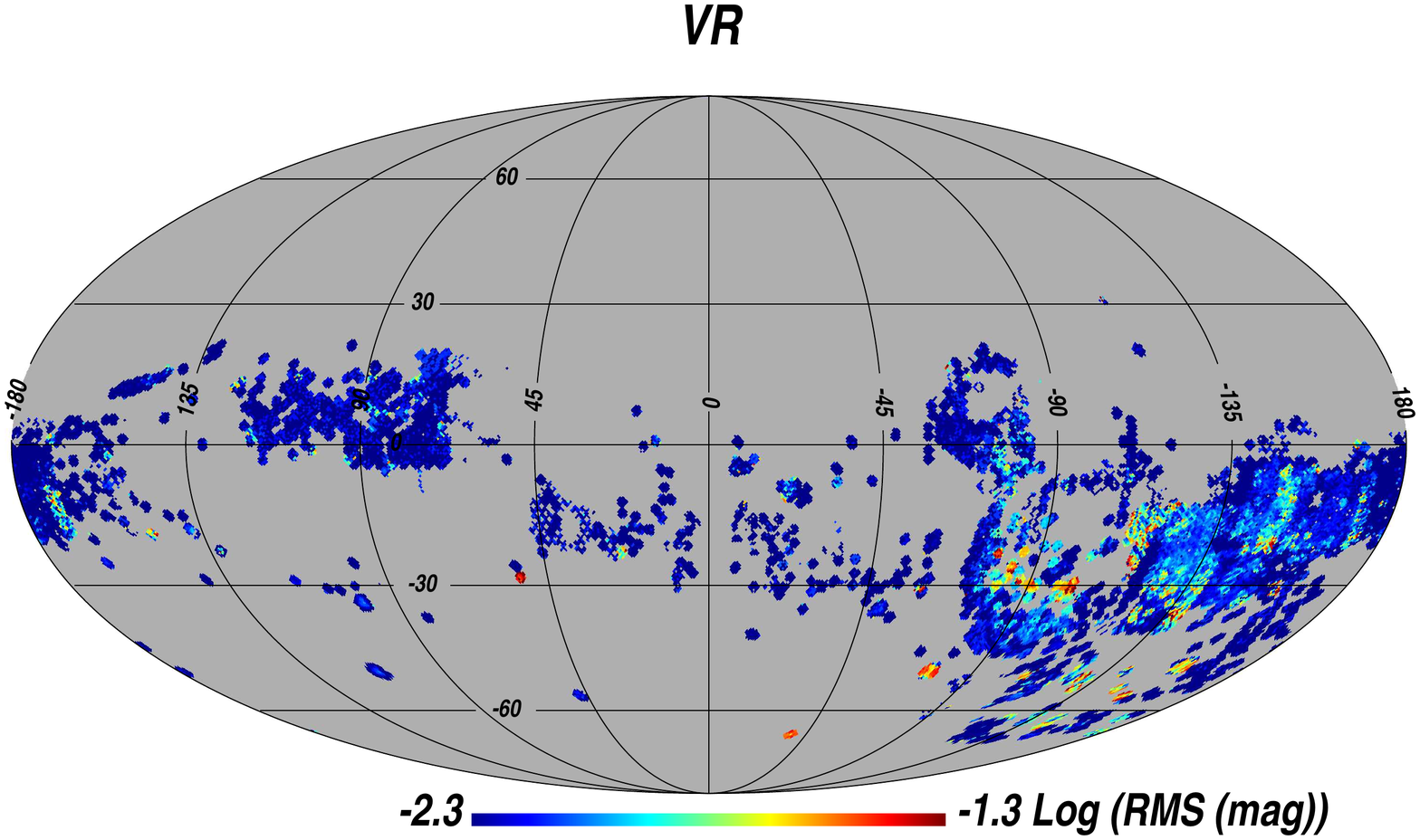}
\end{array}$
\caption{Maps of the NSC DR2 photometric rms of bright stars (with more than two measurements) for the seven $u, g, r, i, z, Y$ and {\em VR} bands on a logarithmic scale in equatorial Aitoff projection.}
\label{fig_photscatter_maps}
\end{figure*}
\end{center}

\begin{center}
\begin{figure*}[!ht]
$\begin{array}{cc}
\includegraphics[trim={0cm 4.9cm 2cm 1cm},clip,width=0.48\hsize,angle=0]{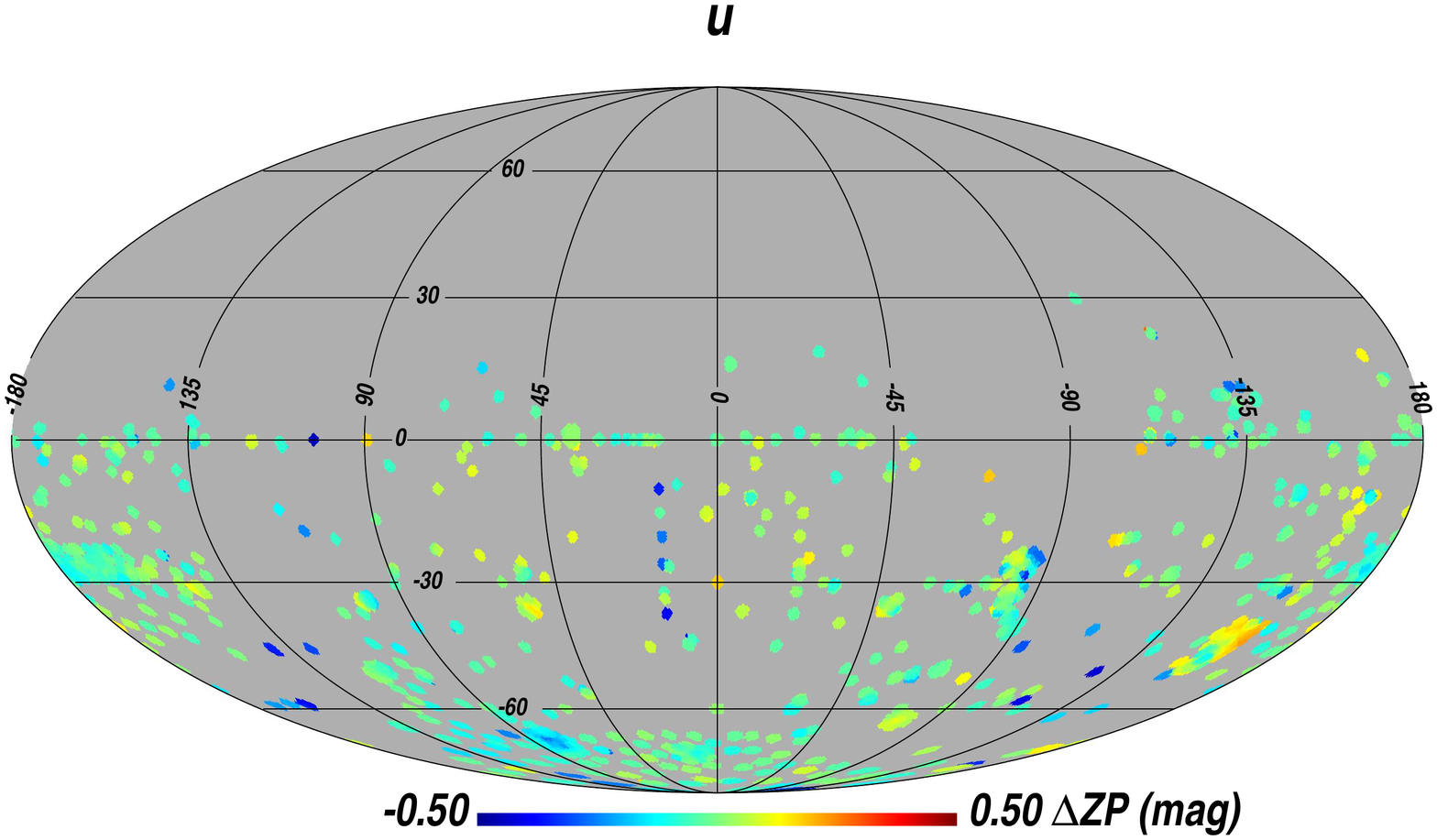}
\includegraphics[trim={0cm 4.9cm 2cm 1cm},clip,width=0.48\hsize,angle=0]{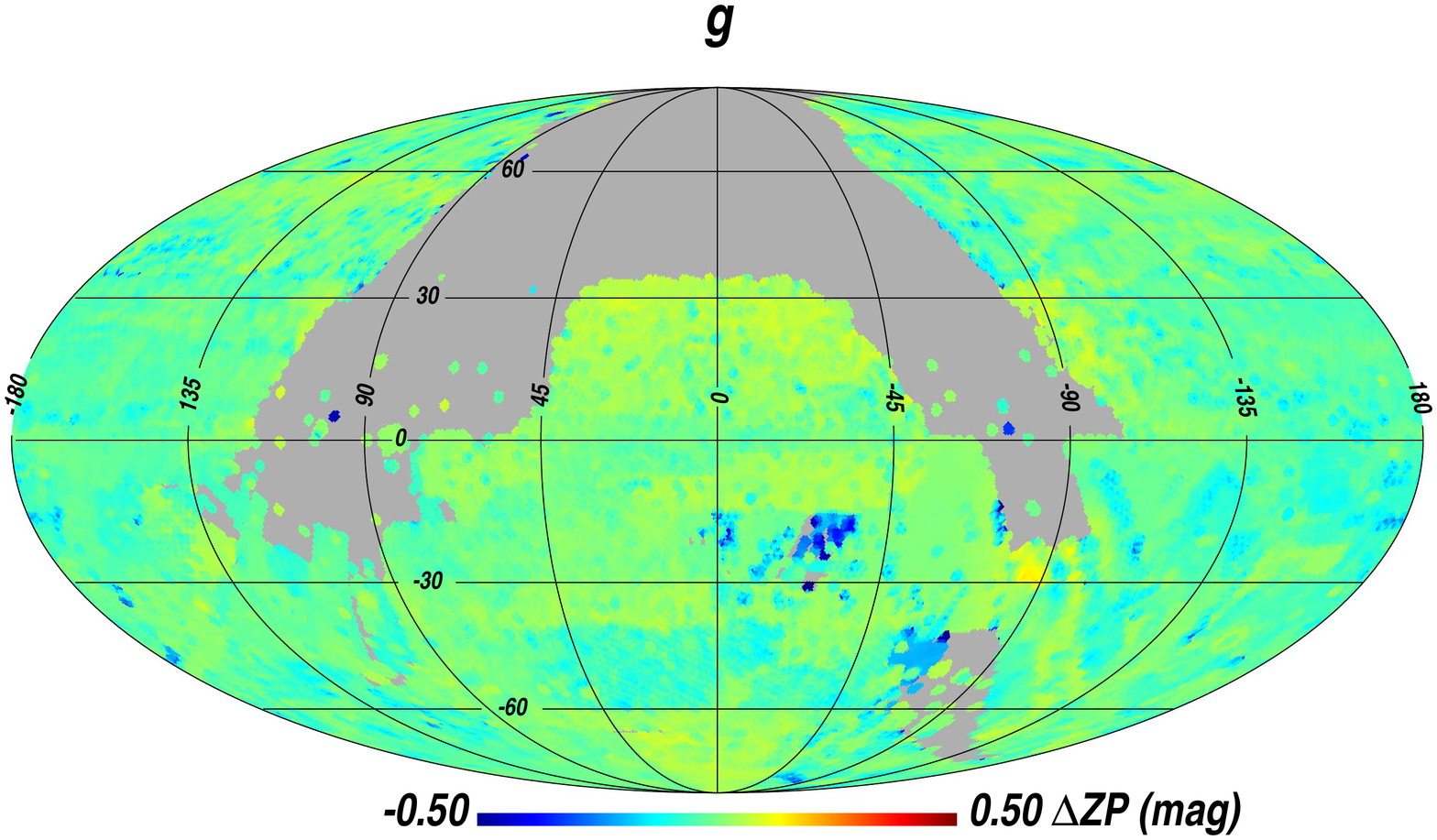} \\
\includegraphics[trim={0cm 4.9cm 2cm 1cm},clip,width=0.48\hsize,angle=0]{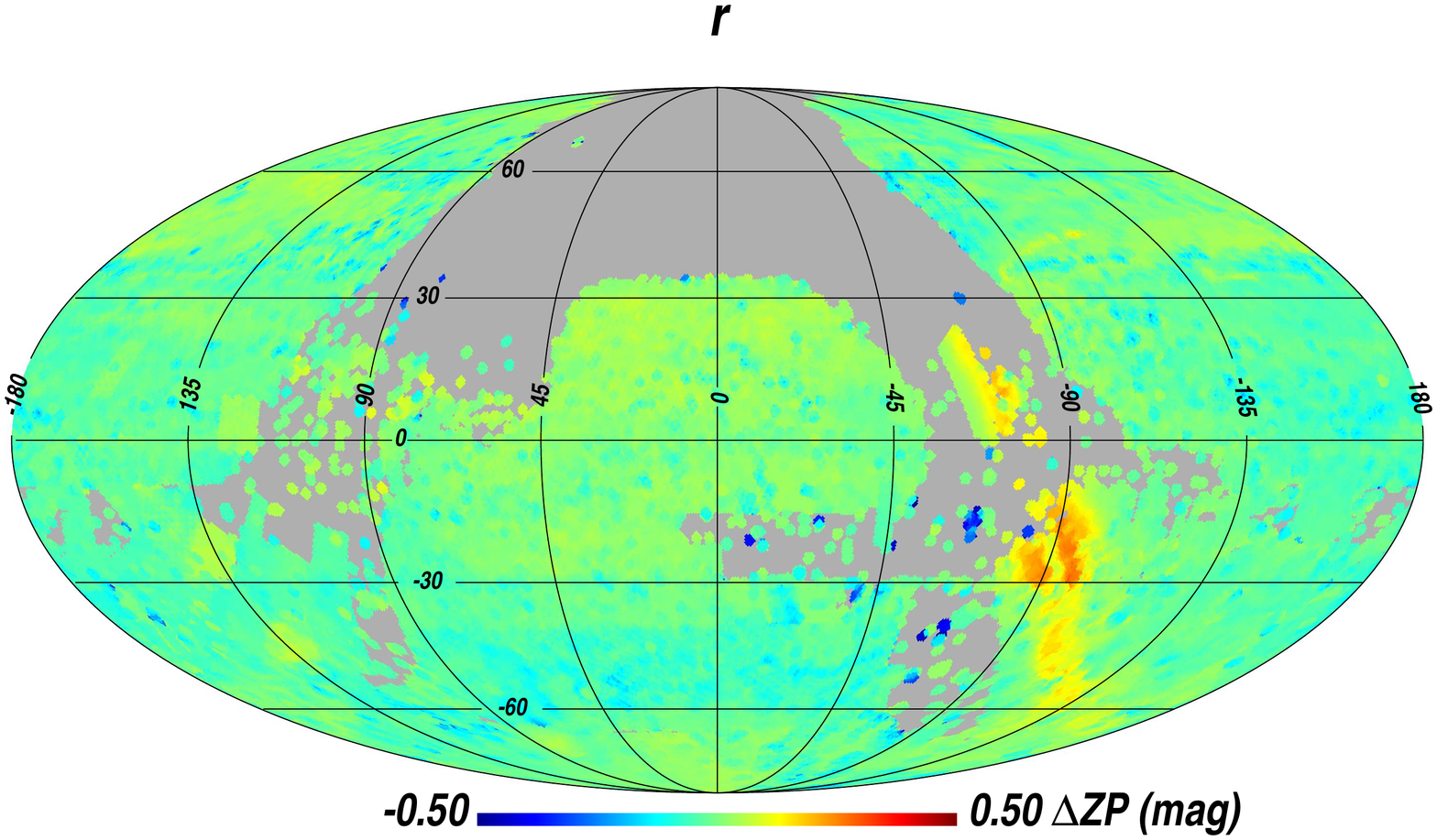}
\includegraphics[trim={0cm 4.9cm 2cm 1cm},clip,width=0.48\hsize,angle=0]{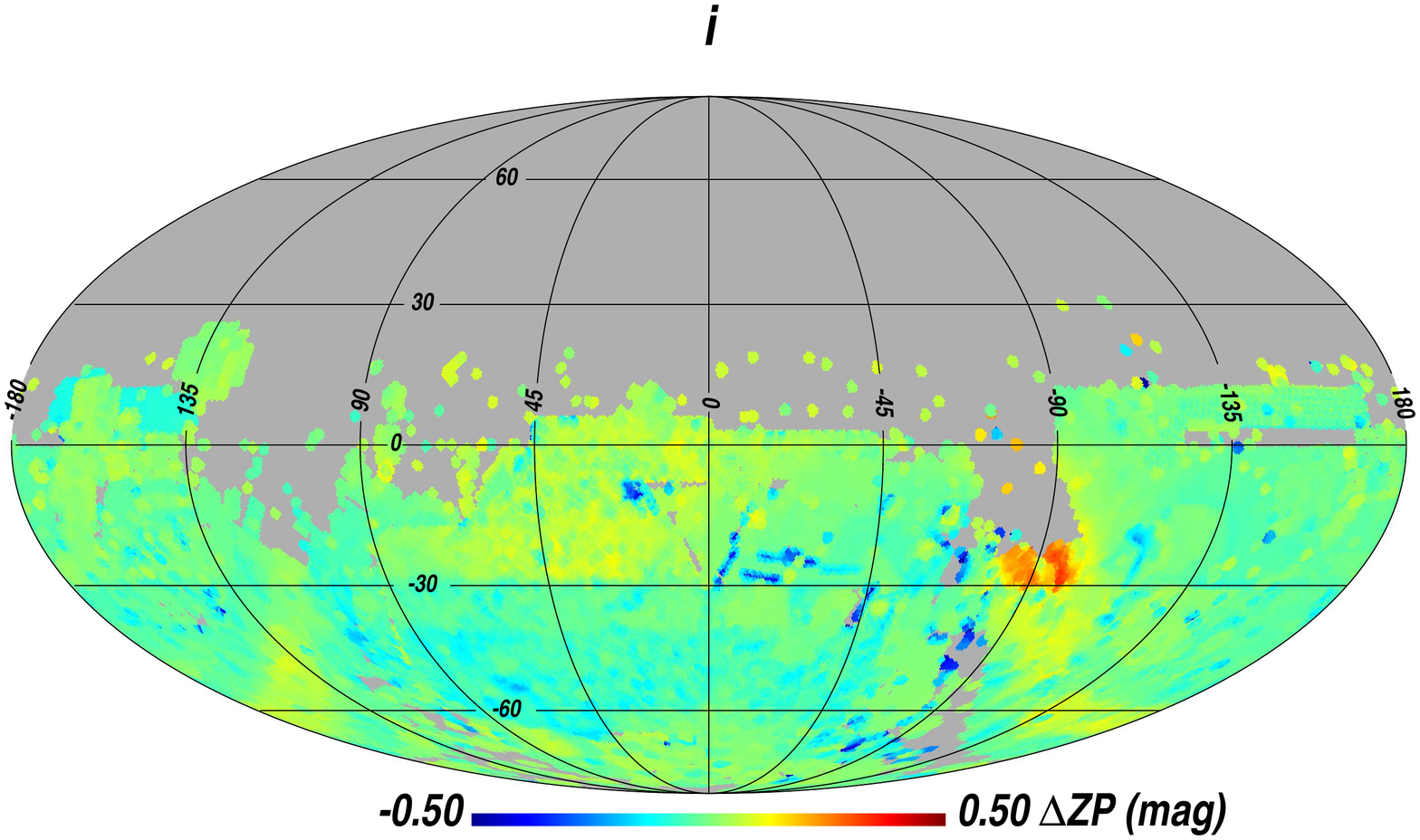} \\
\includegraphics[trim={0cm 4.9cm 2cm 1cm},clip,width=0.48\hsize,angle=0]{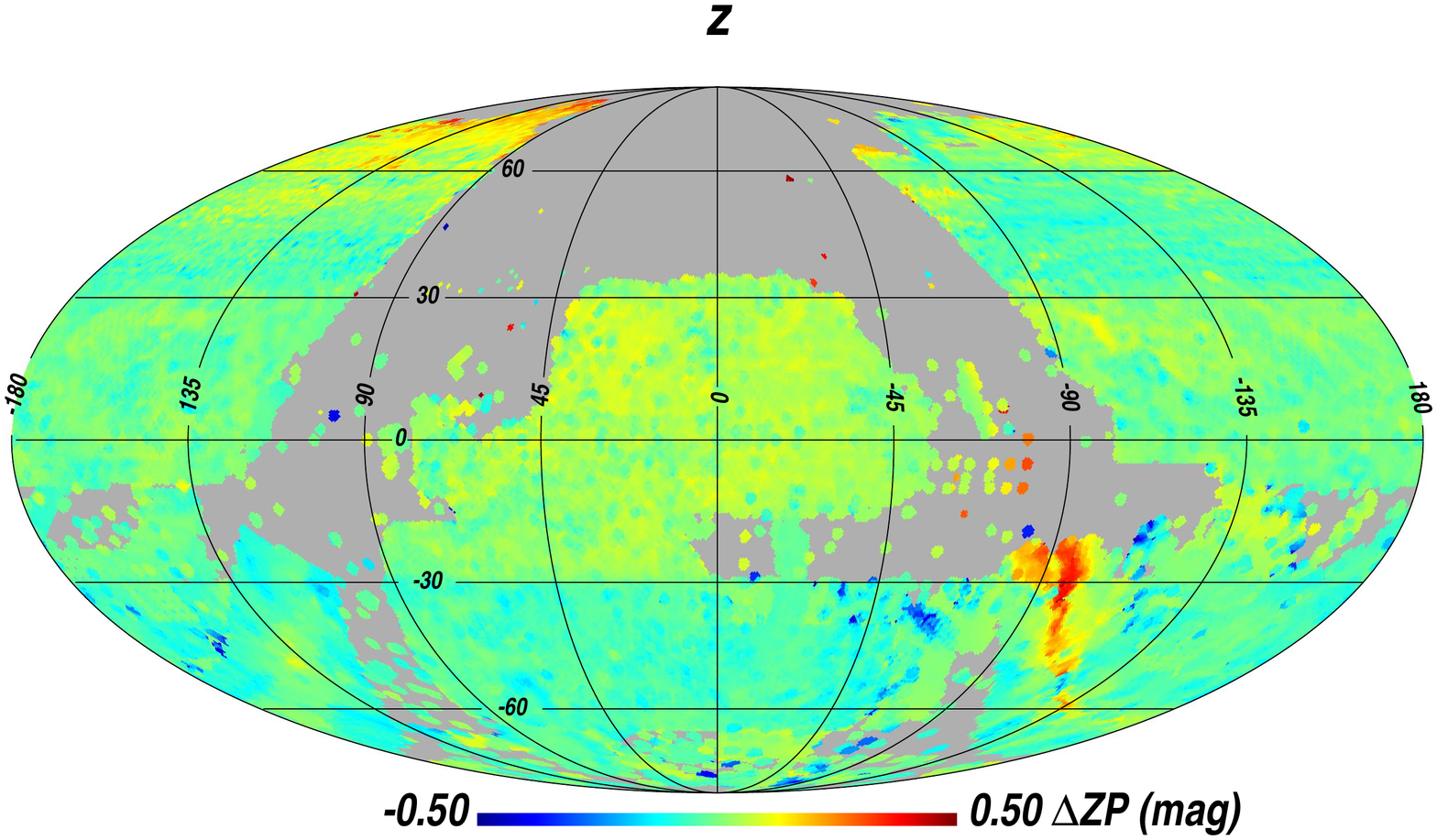}
\includegraphics[trim={0cm 4.9cm 2cm 1cm},clip,width=0.48\hsize,angle=0]{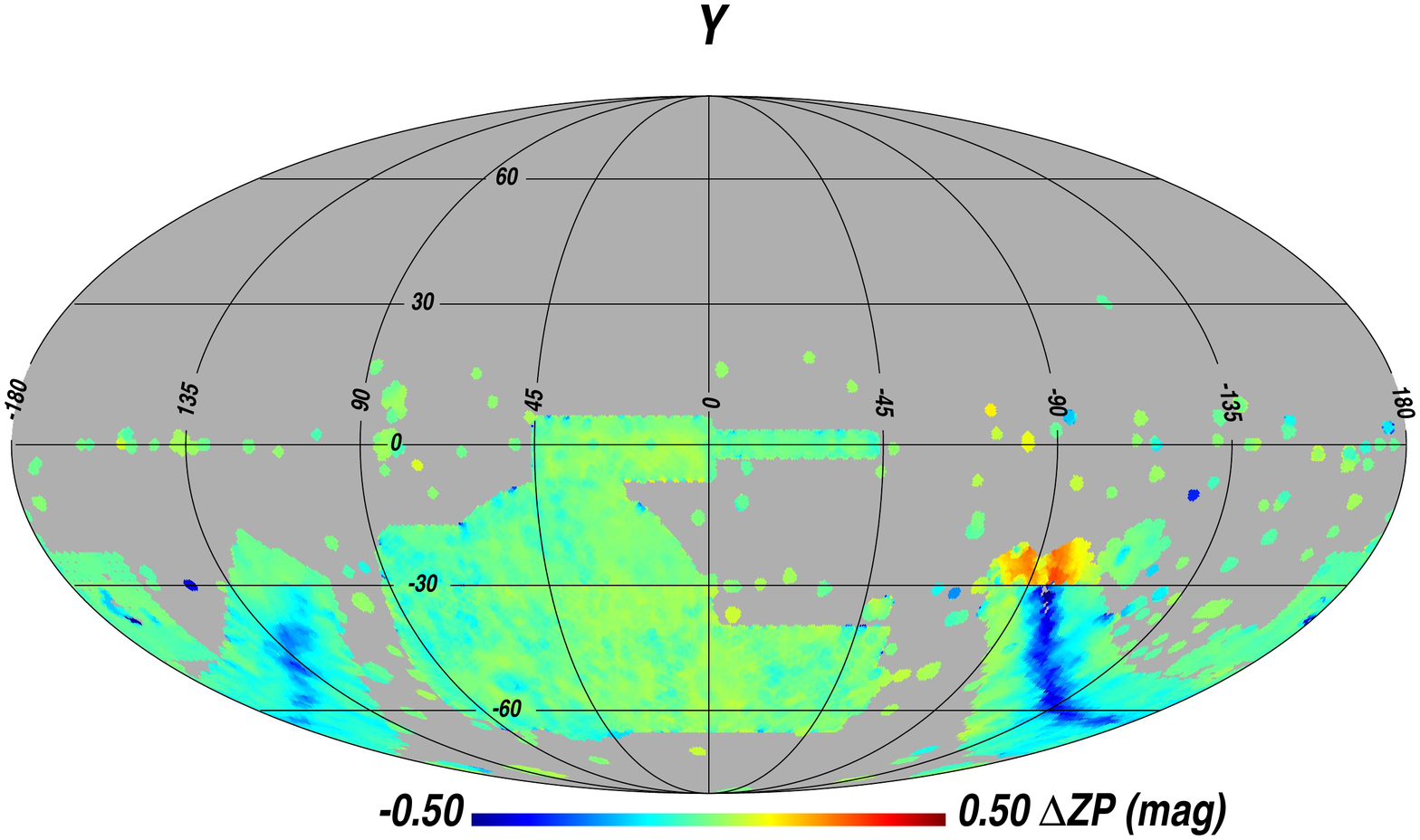} \\
\includegraphics[trim={0cm 4.9cm 2cm 1cm},clip,width=0.48\hsize,angle=0]{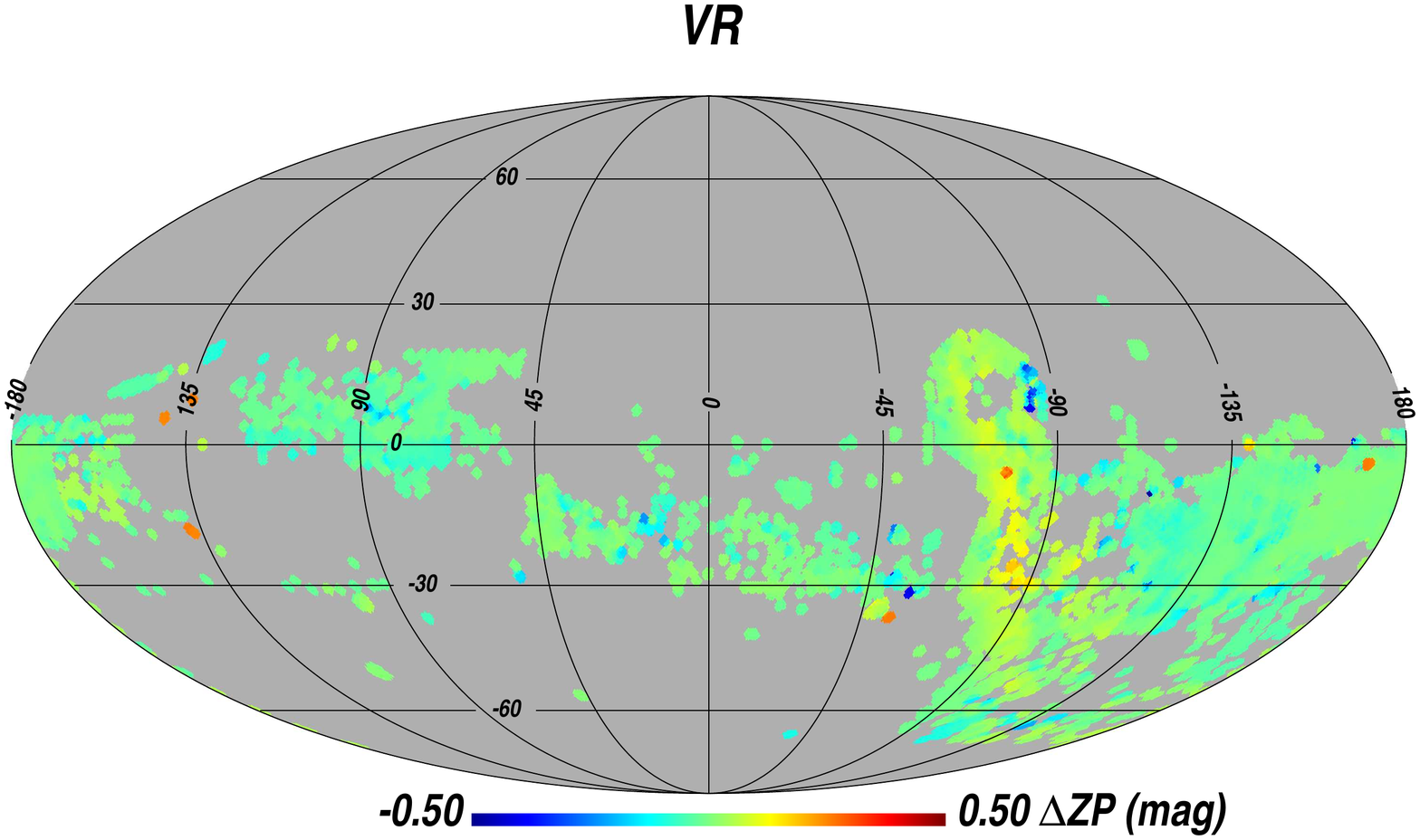}
\end{array}$
\caption{Maps of the mean photometric zero point in each HEALPix relative to the mean across all exposures in a given band (in equatorial Aitoff projection).  The airmass-dependent extinction effects per exposure and long-term temporal variations in the zero points have been removed.}
\label{fig_zeropoint_maps}
\end{figure*}
\end{center}

\begin{center}
\begin{figure*}[!ht]
$\begin{array}{cc}
\includegraphics[trim={1.7cm 4.9cm 2cm 1cm},clip,width=0.49\hsize,angle=0]{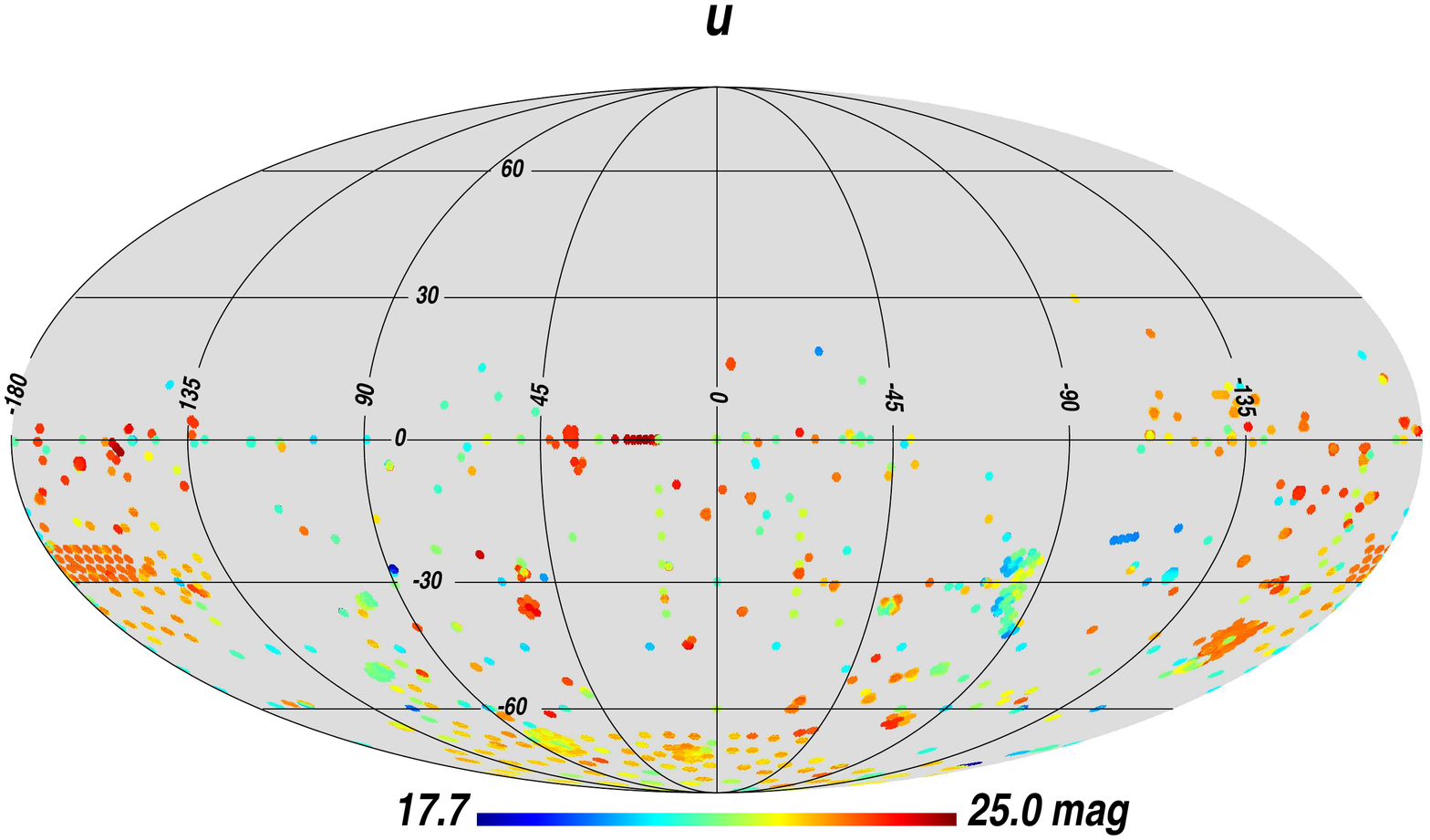}
\includegraphics[trim={1.7cm 4.9cm 2cm 1cm},clip,width=0.49\hsize,angle=0]{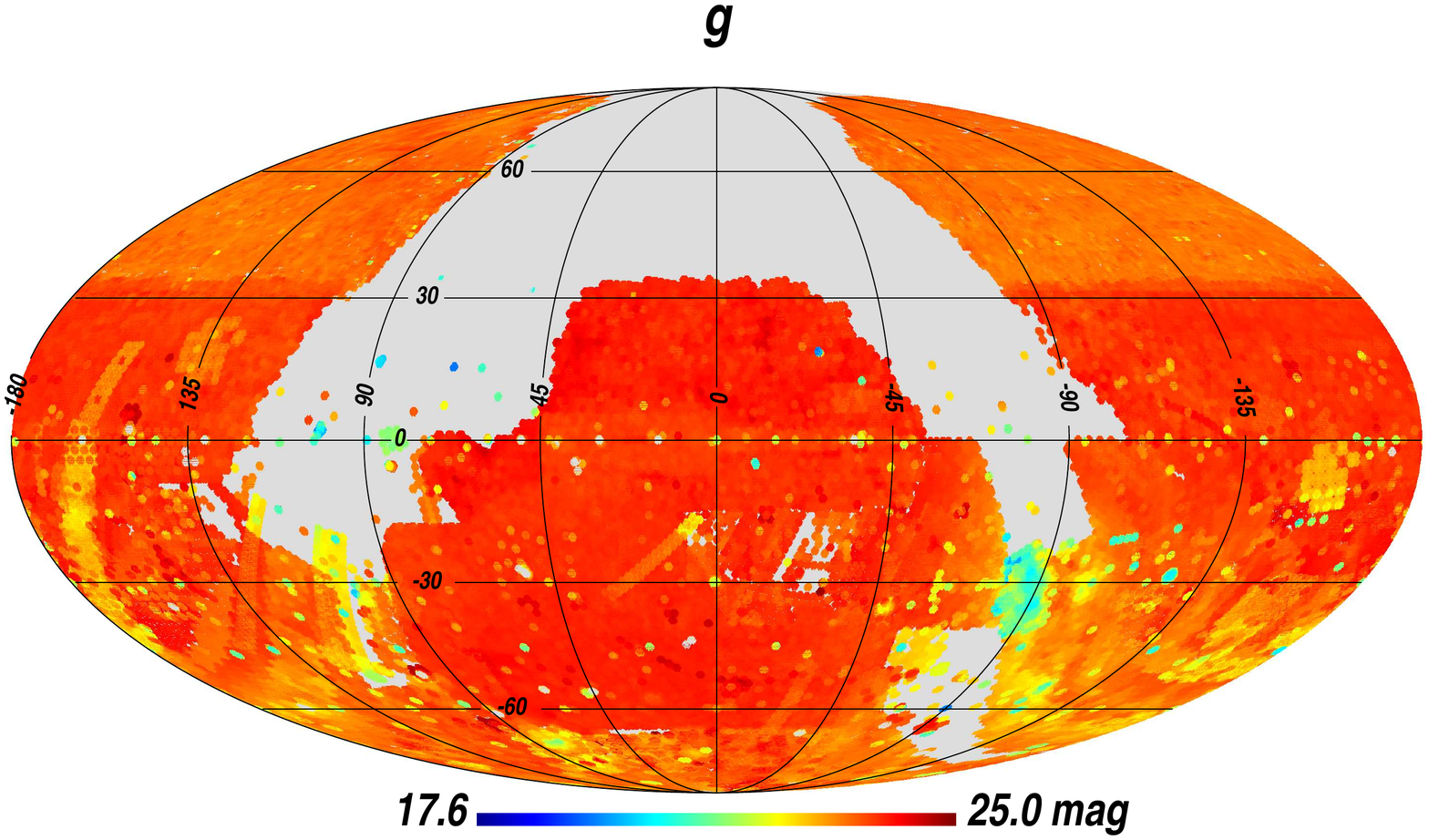} \\
\includegraphics[trim={1.7cm 4.9cm 2cm 1cm},clip,width=0.49\hsize,angle=0]{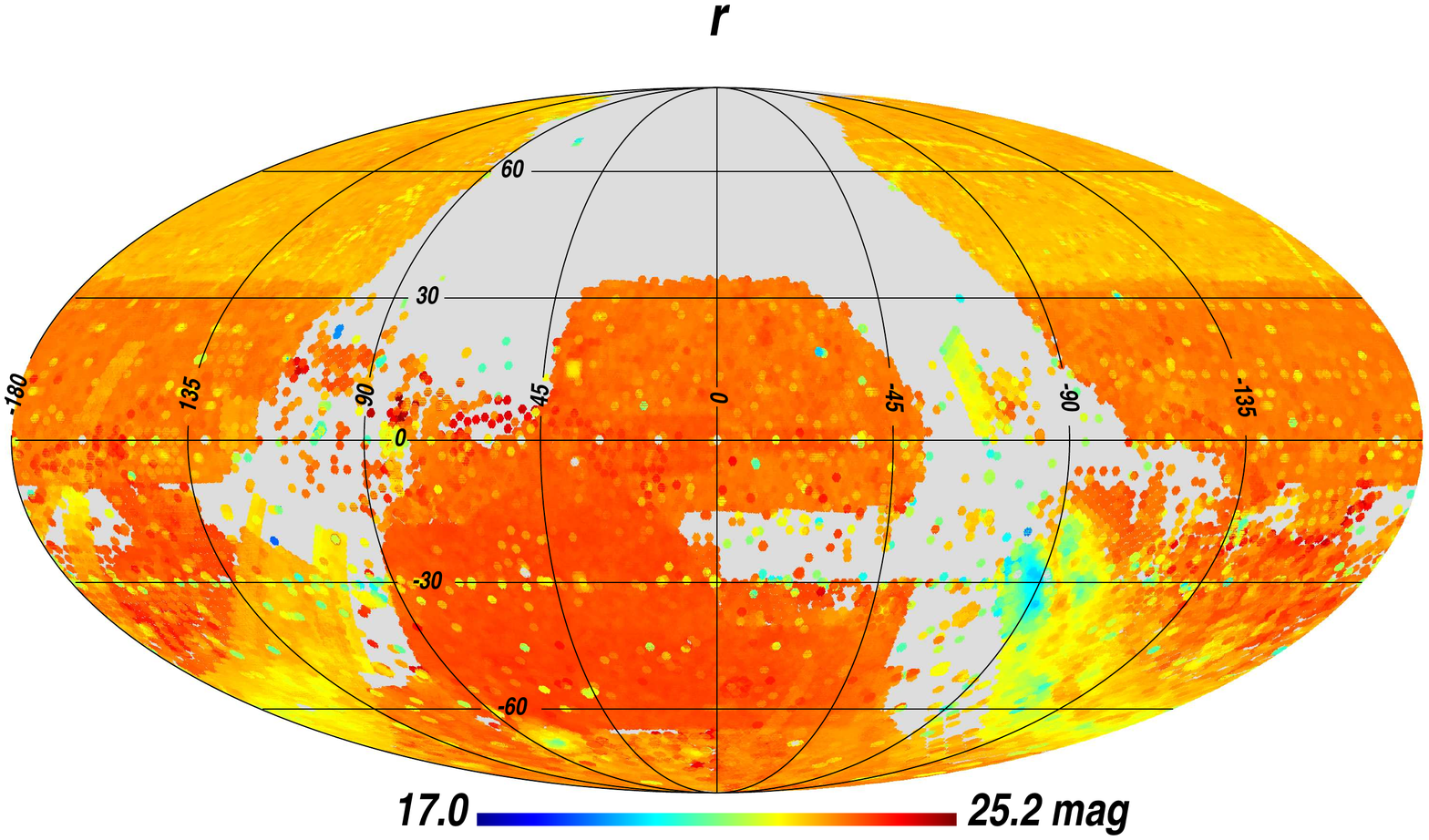}
\includegraphics[trim={1.7cm 4.9cm 2cm 1cm},clip,width=0.49\hsize,angle=0]{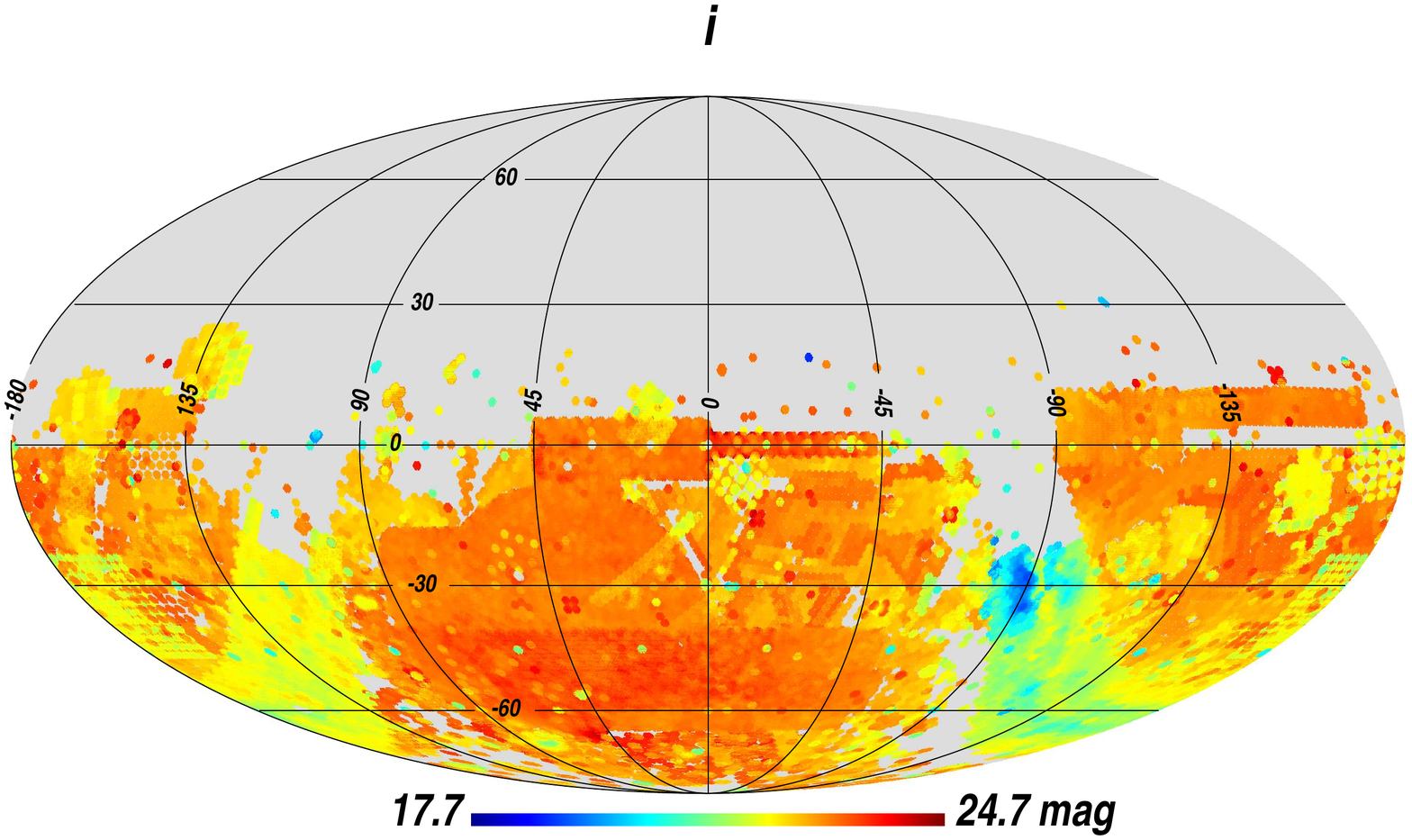} \\
\includegraphics[trim={1.7cm 4.9cm 2cm 1cm},clip,width=0.49\hsize,angle=0]{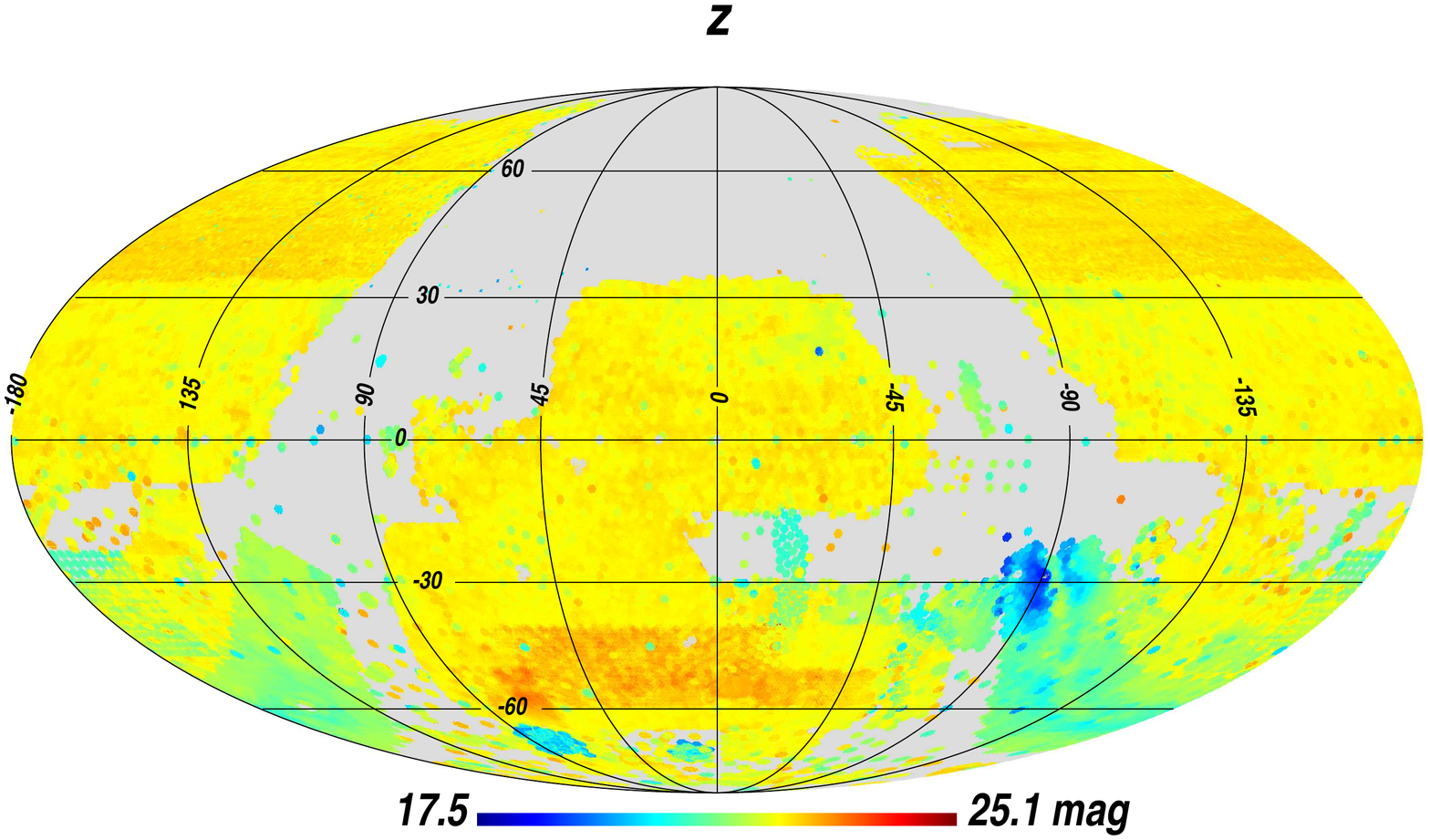}
\includegraphics[trim={1.7cm 4.9cm 2cm 1cm},clip,width=0.49\hsize,angle=0]{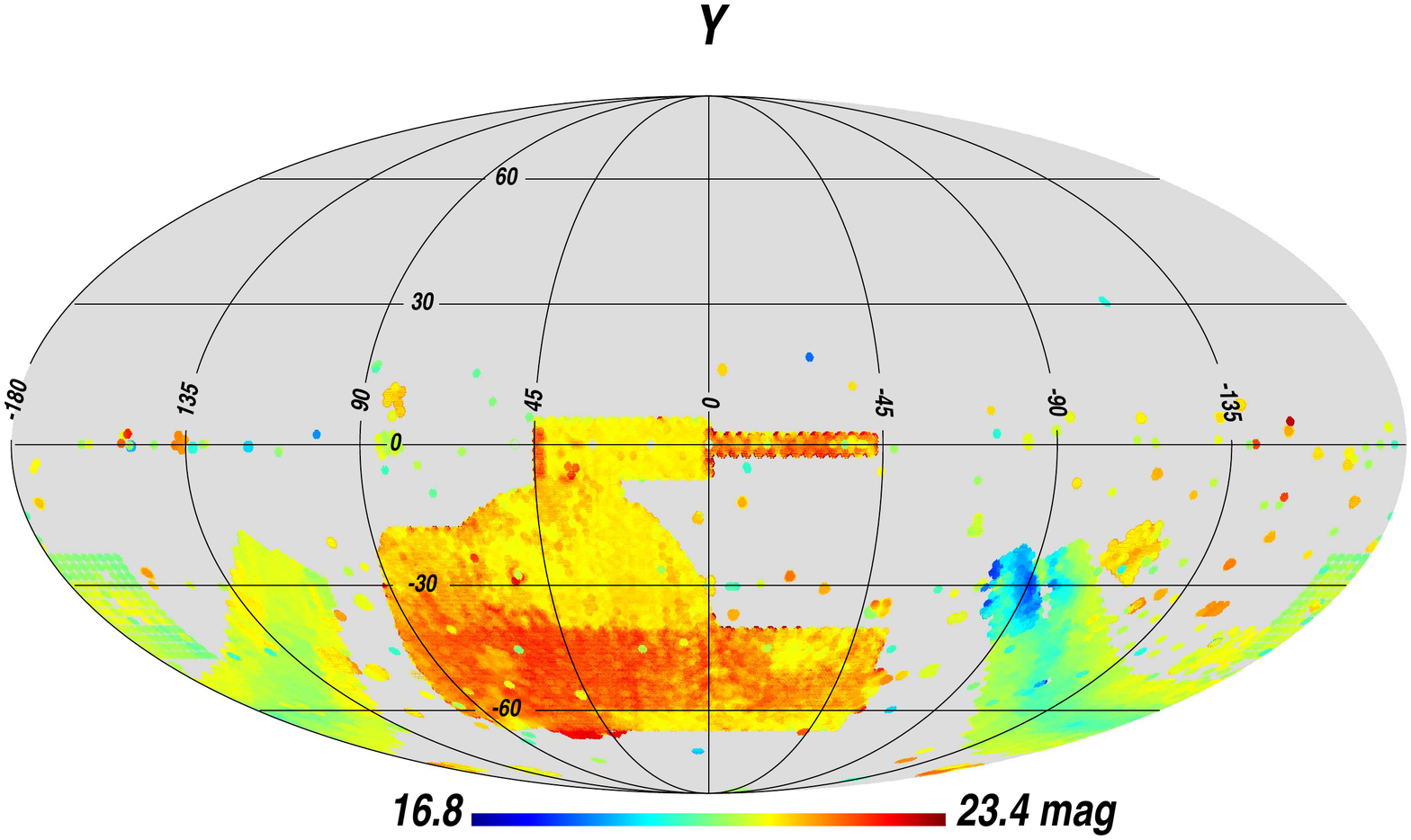} \\
\includegraphics[trim={1.7cm 4.9cm 2cm 1cm},clip,width=0.49\hsize,angle=0]{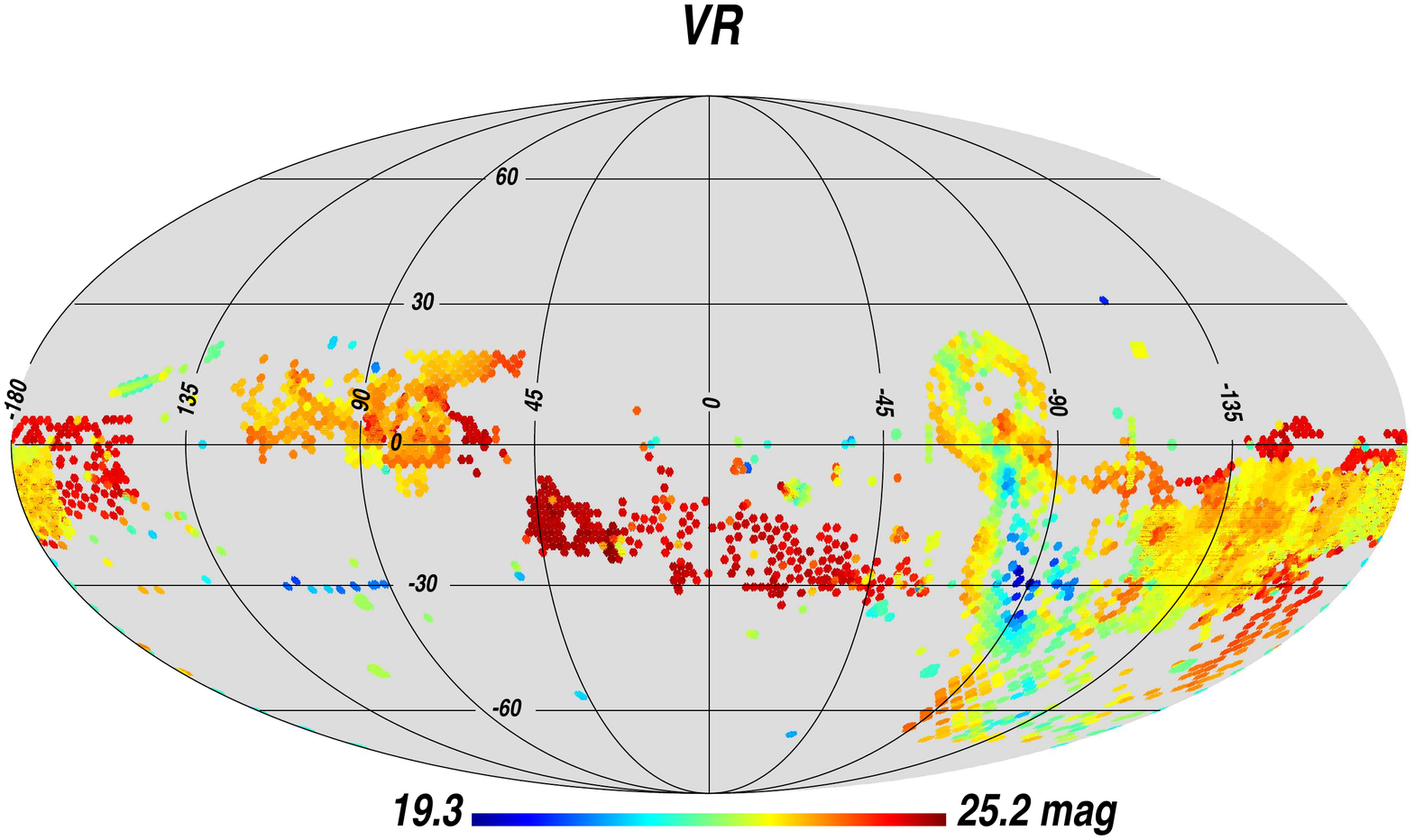}
\end{array}$
\caption{Depth maps (95th percentile) for all seven $u, g, r, i, z, Y$ and {\em VR} bands in equatorial Aitoff projection.}
\label{fig_depths}
\end{figure*}
\end{center}

\subsubsection{Photometry}
\label{subsubsec:photometry}

While the PS1 catalog made it fairly straightforward to calibrate the majority ($grizY$ band) of the northern photometry in NSC DR2, the lack of large-scale photometric surveys and publicly available catalogs made it somewhat challenging to photometrically calibrate the southern data.  We, therefore, relied on ``model magnitudes'', which are linear combinations of photometric measurements from catalogs such as 2MASS \citep{Skrutskie2006} and APASS \citep{Henden2015} that best approximated PS1 $grizY$ and SMASH $u$-band photometry.  Fortunately, the release of Skymapper DR1 and the ATLAS-Refcat2 (which combine data from PS1, Skymapper DR1, ATLAS and other catalogs) made it easier to calibrate southern data in NSC DR2 and decreased our reliance on the 2MASS-APASS-derived model magnitudes.  

For exposures in $grizY$ bands with $\delta$$>$$-$29\degr, zeropoints were derived with PS1 and stars with 0.0$\leq$$(g_{\rm PS1}-i_{\rm PS1})$$\leq$3.0.  For the southern ($\delta$$<$$-$29\degr) exposures in the $griz$ bands, zeropoints were derived using ATLAS-Refcat2 stars with 0.20$\lesssim$$(g_{\rm ATL}-i_{\rm ATL})$$\lesssim$0.80.  For $u$-band exposures with $-$90\degr$\leq$$\delta$$<$0\degr, zeropoints were derived using Skymapper DR1 and stars with 0.80$\leq$$(G_{\rm GAIA}-J)_0$$\leq$1.1.  For $VR$-band exposures, we used the average of the $r$-band (PS1 in the north and ATLAS-Refcat2 in the south) and Gaia DR2 $G$ magnitudes and stars with 0.0$\leq$$(g-i)$$\leq$3.0 to derive zeropoints.  Finally, model magnitudes were used for $u$-band exposures with $\delta$$>$0\dgr and $Y$-band exposures with $\delta$$<$-29\dgr (see Table 1).

We improved our extinction measurements in high extinction regions by using the Rayleigh-Jeans Color Excess \citep[RJCE;][]{Majewski2011}, which uses near- and mid-infrared photometry to derive accurate extinction values star-by-star.  In low extinction regions ($|b|$$>$$16$\dgr and $R_{\rm LMC}$$>$$5$\dgr and $R_{\rm SMC}$$>$$4$\dgr and maximum $E(B-V)<0.2$) the SFD \citep{Schlegel1998} reddening value is used (converted to $E(J-K_{\rm s})$ with a factor of 0.453).  In high extinction regions, RJCE reddening values are used with 2MASS near-infrared photometry \citep{Skrutskie2006} and mid-infrared photometry from $Spitzer$, where possible (from GLIMPSE \citet{Benjamin2003} in the Galactic midplane and SAGE \citet{Meixner2006} in the Magellanic Clouds), or AllWISE \citep{Cutri2013}.  The equation used with $Spitzer$ data is:
\begin{equation}
E(J-K_{\rm s}) = 1.377 (H-[4.5\mu]-0.08); 
\end{equation}
and with AllWISE data is:
\begin{equation}
E(J-K_{\rm s}) = 1.377 (H-W2-0.05)
\end{equation}


Figure \ref{fig_photscatter_maps} shows the rms of photometric measurements of bright stars across the sky in each of the seven bands.  The photometric precision is $\lesssim$10 mmag in all bands (except for $u$-band) across most of the sky.  As in NSC DR1, the photometric scatter is higher in crowded regions like the Galactic midplane and the centers of the LMC and SMC, reaching values of $\sim$50 mmag.  The precision should improve in these regions once PSF photometry is used for measurement.  Figure \ref{fig_zeropoint_maps} shows the maps of mean NSC DR2 zero points with airmass-dependent extinction effects and long-term temporal variations removed.  Table 2
gives statistics on the zero point rms for each band.  Overall, the zeropoints are quite spatially smooth except for crowded regions.  Since we ``absorb'' any aperture correction term into the zero point value, it is not unexpected for this correction to change in crowded versus uncrowded regions and show up in these mean zero point maps. In addition, the jump in the mean zero point of $Y$ in the Galactic midplane across the $\delta$=$-$29 boundary, going from PS1 as the reference in the north to model magnitudes with 2MASS photometry in the south, suggests a systematic issue related to extinction, crowding or aperture corrections in one or both surveys (e.g., PS1 and 2MASS).


\begin{center}
\begin{figure*}[ht]
\includegraphics[trim={1cm 5cm 1cm 1cm},clip,width=1.0\hsize]{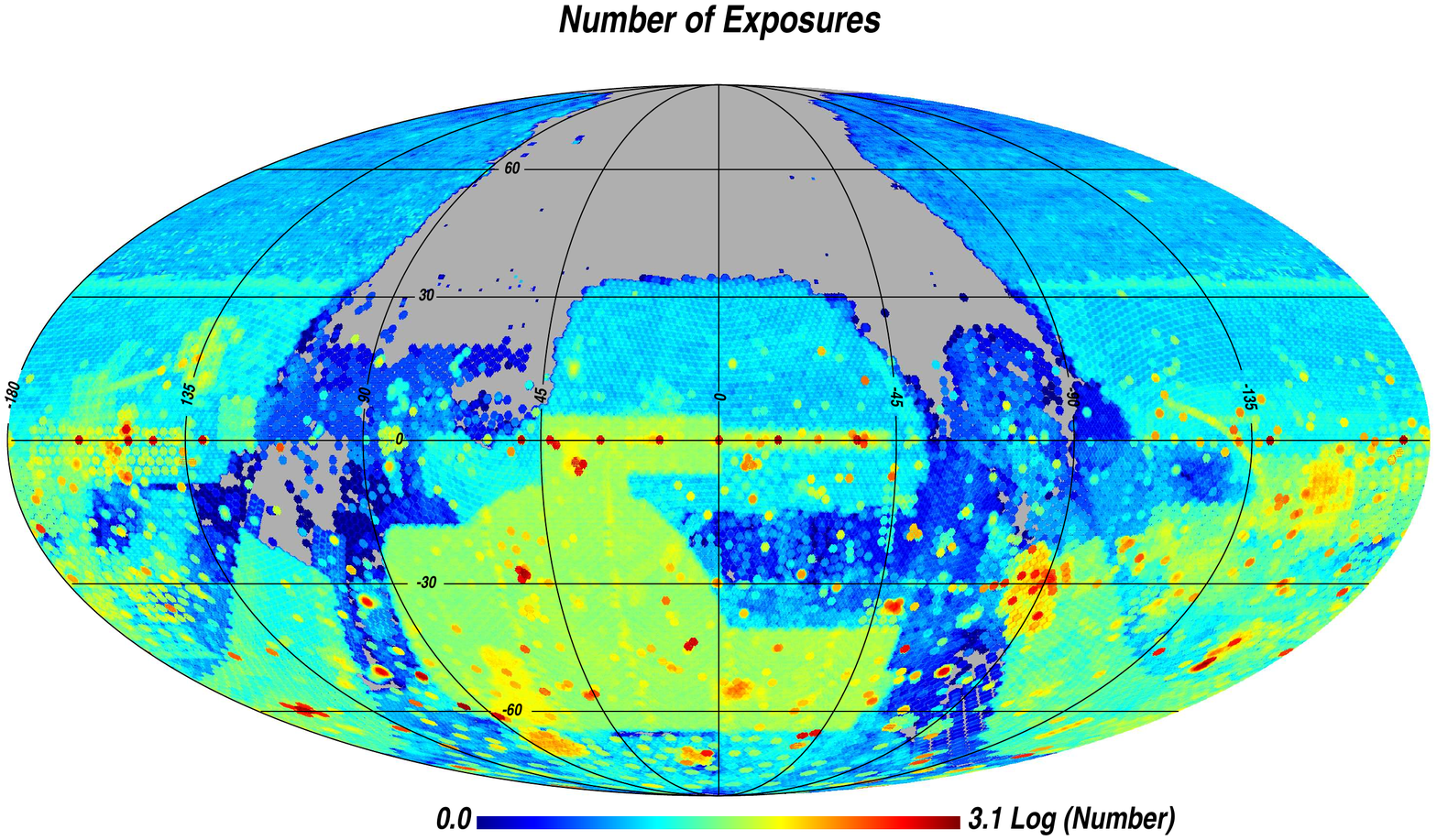}
\caption{Number of NSC exposures on a logarithmic scale in equatorial coordinates.}
\label{fig_nexp}
\end{figure*}
\end{center}

\begin{center}
\begin{deluxetable*}{lc}
\tablecaption{NSC DR2 Model Magnitude Equations}
\tablecolumns{2}
\tablehead{
  \colhead{Model Magnitude} & \colhead{Color Range}
}
\startdata
\vspace{0.1cm} 
$u$ = 0.2301$\times$$NUV_{\rm GALEX}$ + 0.7616$\times$$G_{\rm Gaia}$ + 0.4937$\times$$(G-J)_0$ + 0.8327$\times$$E(J-K_{\rm s})$ + 0.1344    &    0.8$\le$$(G-J)_0$$\le$1.1 \\
\vspace{0.1cm} 
$Y$ = $J$ + 0.54482$\times$$(J-K_{\rm s})_0$ + 0.422$\times$$E(J-K_{\rm s})$ + 0.66338   &       0.4$\le$$(J-K_{\rm s})_0$$\le$0.7 \\
\hline \\
\vspace{0.1cm} 
$(G-J)_0$ = $G_{\rm Gaia}$ $-$ $J$ $-$ 3.27$\times$$E(J-K_{\rm s})$ & \\
$(J-K_{\rm s})_0$ = $J$ $-$ $K_{\rm s}$ $-$ $E(J-K_{\rm s})$ & 
\enddata
\label{table_modelmags}
\end{deluxetable*}
\end{center}

\begin{center}
\begin{deluxetable}{lcccc}
\tablecaption{Zero Point Statistics}
\tablecolumns{5}
\tablehead{
  \colhead{Filter} & \colhead{$\delta$ Range} & \colhead{Median} & \colhead{Median}
  & \colhead{Median} \\
    &  & \colhead{ZP RMS} & \colhead{ZP Error} & \colhead{N$_{\rm reference}$}
}
\startdata
$u$  &  all   &  0.070 &  0.0070  &     552 \\
$g$  & $>-29$ &  0.039 &  0.0005  &    5717 \\
$g$  & $<-29$ &  0.038 &  0.0007  &    3411 \\
$r$  & $>-29$ &  0.036 &  0.0008  &    7290 \\
$r$  & $<-29$ &  0.040 &  0.0010  &    4746 \\
$i$  & $>-29$ &  0.038 &  0.0007  &    9712 \\
$i$  & $<-29$ &  0.056 &  0.0012  &    3473 \\
$z$  & $>-29$ &  0.045 &  0.0018  &    1683 \\
$z$  & $<-29$ &  0.064 &  0.0010  &    4786 \\
$Y$  & $>-29$ &  0.059 &  0.0008  &    7348 \\
$Y$  & $<-29$ &  0.030 &  0.0016  &    1267 \\
{\em VR}  & $>-29$ &  0.030 &  0.0002  &  15024 \\
{\em VR}  & $<-29$ &  0.021 &  0.0003  &  6575
\enddata
\label{table_zpstats}
\end{deluxetable}
\end{center}

\begin{center}
\begin{figure*}[ht]
\includegraphics[width=1.0\hsize,angle=0]{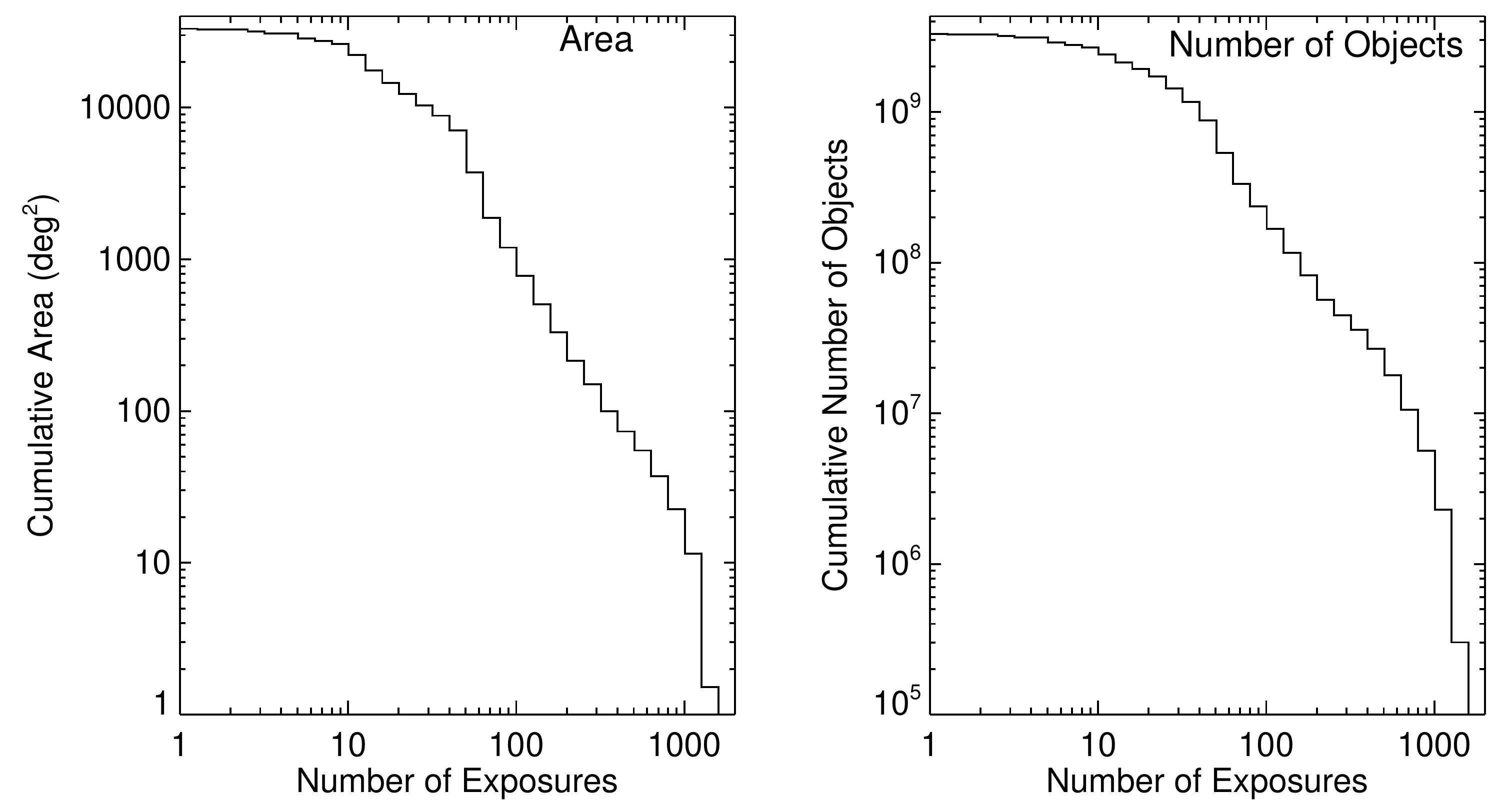}
\caption{Cumulative histogram of ({\em Left}) area and ({\em Right}) number of objects with numbers of exposures greater than some value.}
\label{fig_nexp_cumhist}
\end{figure*}
\end{center}

\subsection{Combination} \label{subsec:combine}

The final step in the NSC processing is ``combination'' in which the measurements from multiple exposures are spatially cross-matched and average properties are calculated for each unique object.

\subsubsection{Quality Cuts}
\label{subsubsec:combineqacuts}


Before the combination process, we first apply quality cuts to the exposures, selecting only data satisfying the following:
\begin{enumerate}
\item public, as of 2019-10-17;
\item all chips astrometrically calibrated (using Gaia DR2) in NSC calibration step;
\item median $\alpha$/$\delta$ RMS across all chips $\leq$0.15\arcsec;
\item seeing FWHM $\leq$2\arcsec;
\item zero point (corrected for airmass extinction) within 0.5 mag of the temporally-smoothed\footnote{The zero points were B-spline smoothed over $\approx$200 nights to track system throughput variations.} zero point for that band;
\item zero point uncertainty $\leq$0.05 mag;
\item number of photometric reference stars $\geq$5 (per CCD);
\item spatial variation (RMS across chips) of zero point $\leq$0.15 mag ($|b|$$>$10\degr) or $\leq$0.55 mag ($|b|$$\leq$10\degr)
(only for DECam with number of chips with well-measured chip-level zero points $>$5);
\item not in a survey's bad exposure list (currently only for the Legacy Surveys and SMASH data).
\end{enumerate}
 
The same quality cuts used in NSC DR1 are applied to the individual measurements.  We only use measurements:
\begin{enumerate}
\item with no CP mask flags set;
\item with no SExtractor object or aperture truncation flags;
\item not detected on the bad amplifier of DECam CCDNUM 31 (if MJD$>$56,600 or big background jump between amplifiers);
\item with S/N$\geq$5.
\end{enumerate}

\subsubsection{Grouping Measurements}
\label{subsubsec:groupmeas}


In NSC DR1, we used a ``sequential clustering'' algorithm to cluster source measurements into objects. Sources were successively crossmatched (with a 0.5\arcsec~matching radius) to existing ``objects" or were added as new objects if no match was found.  Average properties were calculated in a cumulative fashion as measurements were ``added'' to an object.  While this algorithm was efficient, it did not allow the use of robust statistics (e.g., outlier rejection), the calculation of photometric variability indices, or the ability to detect fast-moving objects.  In NSC DR2, we employed a hybrid spatial clustering algorithm to group measurements into objects.  As in NSC DR1, the HEALPix scheme \citep{Gorski2005} with NSIDE=128 is used to tile the sky into smaller regions to efficiently parallelize the computation during this combination step.  

For a given HEALPix, all measurements passing the above-mentioned quality cuts of chip images overlapping the HEALPix and its neighboring HEALPix are loaded. 
For HEALPix with many measurements (over 1 million), the combination algorithm is performed on smaller HEALPix subregions (up to 64 nside=1024) and the results later merged together.

The two steps of the hybrid spatial clustering algorithm are (1) clustering with DBSCAN \citep[Density-based spatial clustering of applications with noise][]{Ester96} using a small clustering distance to generate object centers, followed by (2) sequential clustering of the leftover measurements using the object centers.  The first step allows the definition of objects themselves (i.e., their central positions) using their spatial coherence which should be roughly on the scale of the median astrometric uncertainty.  Therefore, the eps parameter, the maximum distance that two points within a given cluster can be separated, is set to three times the median astrometric uncertainty or a minimum of 0.3\arcsec; on average eps$\approx$0.4\arcsec.  The minimum number of points to define a cluster is either three or the total number of exposures (if this is $<3$).  The second step is needed because the DBSCAN clustering does not take into account the astrometric uncertainty of individual measurements.  The measurements not clustered in the DBSCAN step are crossmatched to the existing object centers using a crossmatch radius of three times their astrometric uncertainty or a minimum of the DBSCAN eps value.  The crossmatching is done successively, with the leftover measurements from one exposure at a time.  Any measurements not matched to existing objects are added as new objects to the object list.


We then calculate average properties for each object from the calibrated and grouped measurements.  These include flux-weighted mean coordinates, robust proper motions, mean magnitude, uncertainties, RMS, some morphology parameters per band, and mean morphology parameters across all measurements.

\subsubsection{Photometric Variabilty Metrics}
\label{subsubsec:photvar}

\begin{center}
\begin{figure*}[ht]
$\begin{array}{cc}
\includegraphics[width=0.33\hsize,angle=0]{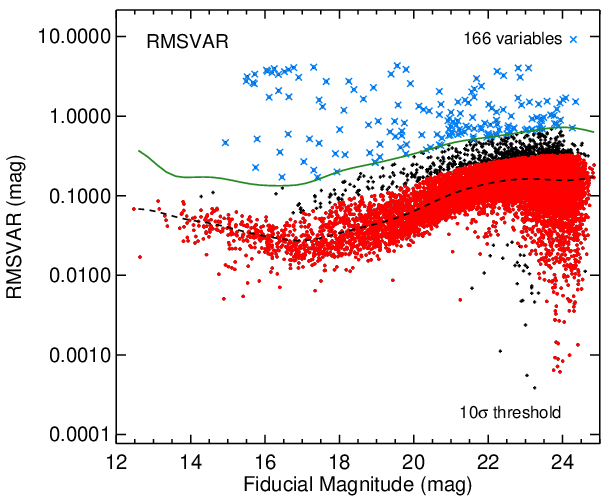}
\includegraphics[width=0.33\hsize,angle=0]{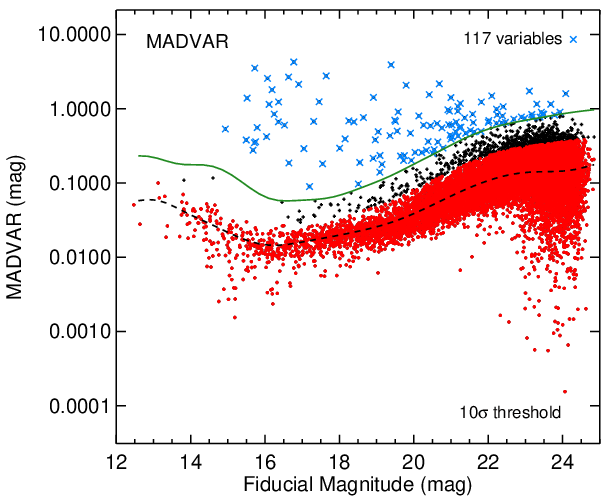}
\includegraphics[width=0.33\hsize,angle=0]{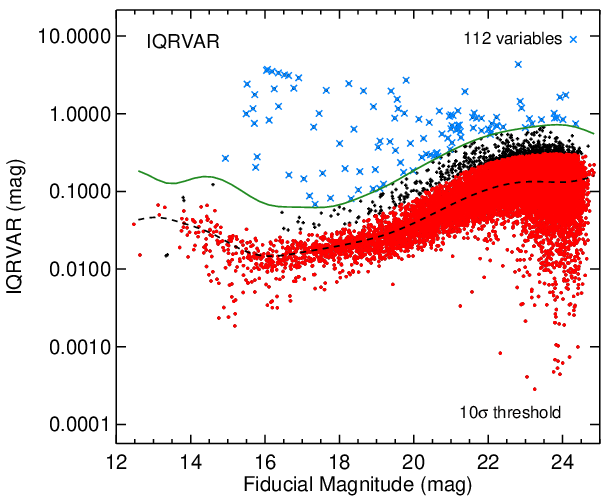} \\
\includegraphics[width=0.33\hsize,angle=0]{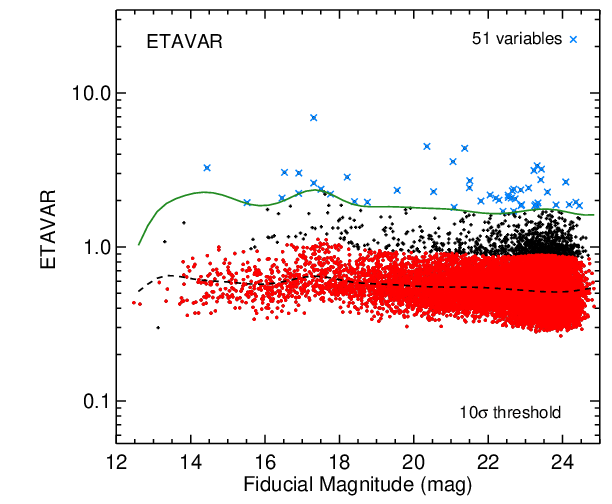}
\includegraphics[width=0.33\hsize,angle=0]{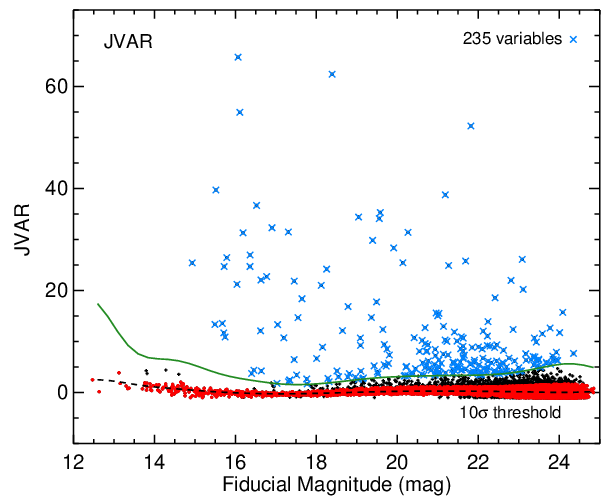}
\includegraphics[width=0.33\hsize,angle=0]{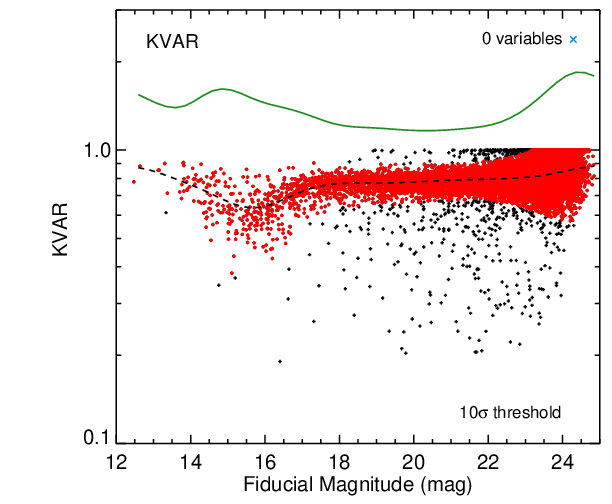} \\
\includegraphics[width=0.33\hsize,angle=0]{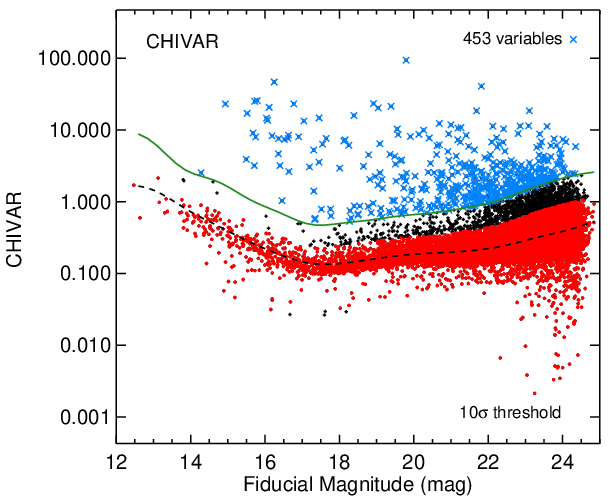}
\includegraphics[width=0.33\hsize,angle=0]{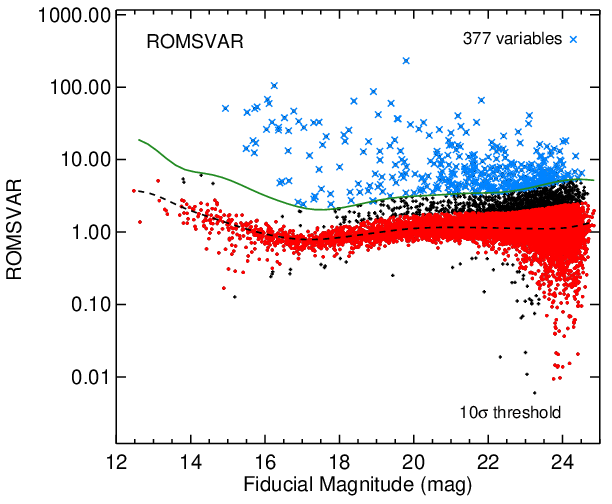}
\end{array}$
\caption{The eight photometric variability indices computed in NSC DR2. RMS, MAD, IQR, von Neumann ratio $\eta$, Stetson's J and K indices, $\chi$, and RoMS.  Each index is shown versus a ``fiducial magnitude'' which is the first band in a prioritized list ($r$, $g$, $i$, $z$, $Y$, $VR$, $u$) that has been observed for a given object.  The filled red circles are objects within 3$\sigma$ of the median as a function magnitude (black dashed line).  The blue $\times$ symbols are objects 10$\sigma$ above the median; this threshold is indicated by the green line.
}
\label{fig_photvar}
\end{figure*}
\end{center}

The new clustering method allows for the calculation of photometric variability indices. We calculate eight variability metrics: RMS, MAD, IQR, von Neumann ratio $\eta$, Stetson's J and K indices, $\chi$, and RoMS.  \citet{Sokolovsky2017} give detailed descriptions and comparisons of these and other metrics and helped guide our work in this area. The metrics we used can be separated into two groups: (1) metrics using only the magnitude residuals (relative to the flux-weighted mean magnitude in each band; i.e., MAD, RMS, IQR, $\eta$), and (2) metrics using both the magnitude residuals and their uncertainties (J, K, $\chi$, RoMS).
Examples of the eight photometric variability indices for one HEALPix are shown in Figure \ref{fig_photvar}.
The photometric variability indices alone are not enough to select photometrically variable objects as the average value of the metric will change with magnitude.  An additional analysis is performed on each HEALPix to calculate the median value of the metric and the robust scatter as a function of magnitude.  We cannot use a single band for this magnitude because not all objects will have data in that band.  Therefore, we construct a ``fiducial magnitude'' which is the first band in the prioritized list [$r$, $g$, $i$, $z$, $Y$, $VR$, $u$] that has been observed for a given object. \autoref{fig_photvar} shows objects within 3$\sigma$ of the median metric value as a function of fiducial magnitude (black dashed line) as filled red circles.  Objects with variability 10$\sigma$ or more above the median are indicated by blue $\times$ symbols (the 10$\sigma$ cutoff is denoted by the green solid line).  We decided to use the MAD variability index for identifying variable sources.  
All objects that are 10$\sigma$ above the median are flagged \texttt{VARIABLE10SIG} (23,270,027 objects).  The offset of each object in units of $sigma$ from the median is reported in the catalog as \texttt{NSIGVAR}, to aid users who desire a different $\sigma$ cutoff.  The use of the NSC DR2 variability information to study variable stars and quasi-stellar objects (QSOs) is discussed in Sections \ref{subsec:variables} and \ref{subsec:qsos}, respectively.

\section{Caveats}
\label{sec:caveats}

Users of the NSC DR2 should be aware of the following caveats.

\begin{center}
\begin{figure}[ht]
\includegraphics[width=1.0\hsize,angle=0]{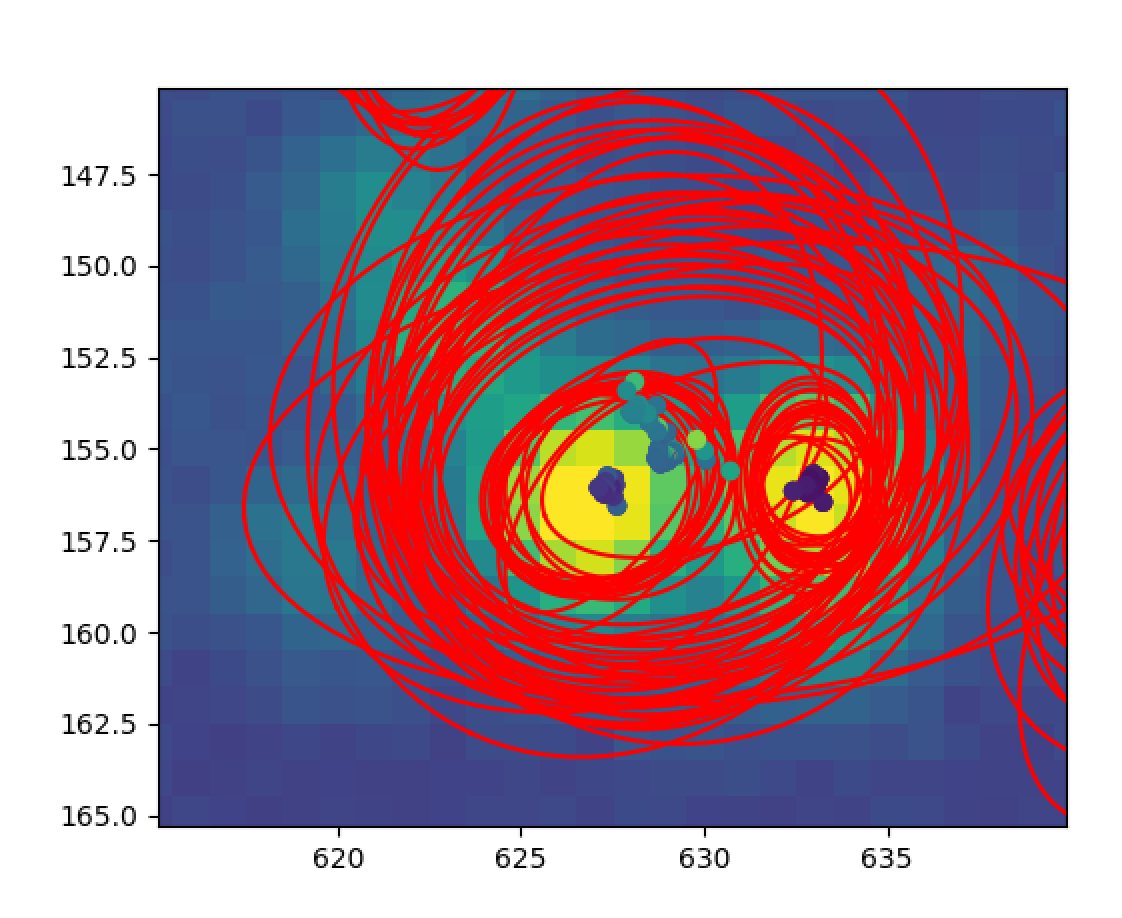}
\caption{Combining measurements of objects taken under different seeing conditions results in source confusion.  The background image is a good-seeing exposure showing two resolved stars.  The dark filled circles are the centers of individual measurements color-coded by their spatial FWHM (darker colors for smaller seeing FWHM).  The red ellipses are the measured shapes of those measurements. The better seeing data results in two sources associated with the two stars, whereas the poor seeing data results in a common source with a larger ellipticity.}
\label{fig_deblending}
\end{figure}
\end{center}

As the observations are taken over a range of observing conditions and instruments, two distinct neighboring objects may be spatially resolved in some exposures but not others.  This causes inherent problem when combining measurements at the catalog-level.   Figure \ref{fig_deblending} shows one example, where the measured object centers cluster into three groups: the individual centers of the two stars from good-seeing exposures, and a position between the two resulting from the poor-seeing exposures where the sources remain confused.   There is no clear-cut ``correct'' way to handle this situation, without a more sophisticated source modeling approach \citep[e.g., {\it Tractor}][]{Tractor}.  For now, we have chosen the simple approach: we have left the three clusters as three separate objects, but flagged the object of the unresolved pair of stars as a \texttt{PARENT}.  This flag is set for any object that contains other objects inside its ellipse footprint defined by its central coordinates and the \texttt{ASEMI}, \texttt{BSEMI}, and \texttt{THETA} shape parameters.

As mentioned in Section \ref{subsec:combine}, the eps parameter was determined independently in each HEALPix based on the measurements and the median astrometric uncertainty within that HEALPix. 
While this was meant to allow the clustering of measurements into objects to be determined by the data itself, it had unforeseen consequences at the boundaries of HEALPix regions.  Each HEALPix region includes measurements in a 10\arcsec\ boundary around it.  Only objects, and their constituent measurements, are included in the HEALPix catalog if the final central position is inside the HEALPix boundary.  If a neighboring HEALPix clusters the measurements at the boundary in the same way, as was done in NSC DR1, then the measurements and objects are appropriately parceled out to their correct HEALPix.  In NSC DR2 the clustering parameter changes slightly from one HEALPix to the next, resulting in rare instances when measurements are either not grouped into an object or grouped to multiple objects.  This mostly happens in very crowded regions such as in the Galactic bulge or the centers of the LMC and SMC.
The NSC DR2 contains 77,273 missing and 9,345 duplicate measurements.  While this is a non-negligible number, it is nonetheless a small fraction of the total 68 billion total measurements.  In a future data release, the DBSCAN clustering parameter will be fixed for all HEALPix.

\section{Description and Achieved Performance of Final Catalog}
\label{sec:performance}



The NSC DR2 covers more than 35,000 square degrees of the sky and catalogs over 3.9 billion unique objects (Fig.\ \ref{fig_bigmap}).  It includes more than 68 billion individual measurements --- twice the number in NSC DR1 --- from 412,116 exposures spanning over 7 years. Most of the sky is covered in multiple bands, with 33,028 square degrees having two bands and 30,860 square degrees having three bands.  Almost 1.9 billion objects have data in three or more bands and can be used to construct color-color diagrams.

Maps of the 95th percentile depths are shown in Figure \ref{fig_depths}. The median depths are 22.6, 23.6, 23.2, 22.8, 22.3, 21.0, 23.3 mag in the $u, g, r, i, z, Y,$ and {\em VR} bands.  The photometric precision (Fig.\ \ref{fig_photscatter_maps}) is $\lesssim$10~mmag is all bands with the exception of the $u$-band, and is fairly uniform across the sky.  Although an effort has been made to improve the photometric calibration in crowded and dusty regions by using more accurate extinction corrections (e.g., the RJCE method) and newer reference catalogs in the southern sky (e.g., Skymapper DR1 and ATLAS-Refcat2) some issues remain.  We advise caution when using the photometry in the very crowded and high extinction regions.

Most of the sky is covered by multiple exposures giving rise to a valuable time-series dataset (Fig.\ \ref{fig_nexp}).  Cumulative histograms of the area and number of objects with a certain number of exposures is shown in Figure \ref{fig_nexp_cumhist}.  Roughly 500 million objects have 30 or more exposures, which should be enough to reliably detect and classify many classes of variable stars (e.g., see Section \ref{subsec:variables}).

\begin{center}
\begin{figure}[t]
\includegraphics[width=1.0\hsize,angle=0]{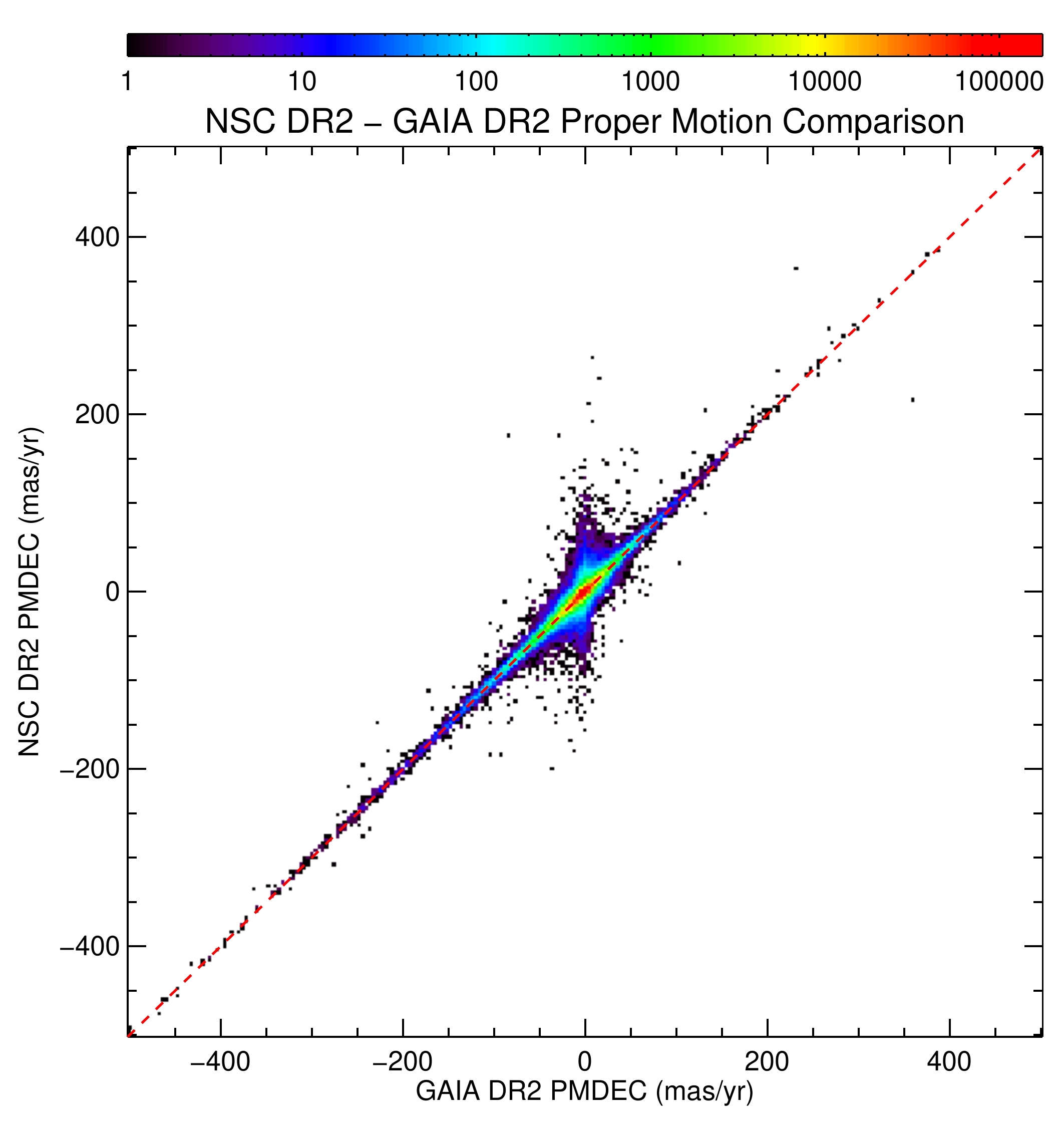}
\caption{Comparison of NSC DR2 proper motion measurements with those from Gaia DR2 for 1,365,136 stars in a 700 degree squared region centered on ($\alpha$,$\delta$)=(45\degr,$-$30\degr). Only stars with proper motion S/N$>$3 or proper motion error $<$3 mas yr$^{-1}$ (in both $\mu_{\alpha}$ and $\mu_{\delta}$) in both catalogs and at least three detections in the NSC and a temporal baseline of 200 days were selected.  The one-to-one line is shown in red. The median offset in $\mu_\alpha$/$\mu_\delta$ is 0.248/0.065 mas yr$^{-1}$ (NSC$-$Gaia) with a robust scatter of 2.45/2.36 mas yr$^{-1}$.}
\label{fig_pmcomparison}
\end{figure}
\end{center}

The large numbers of repeat observations of individual sources also permits reliable estimates of their proper motion. Figure \ref{fig_pmcomparison} shows a comparison of well-measured  NSC DR2 proper motions (S/N$>$3 or a proper motion error $<$3 mas yr$^{-1}$ in both $\mu_{\alpha}$ and  $\mu_{\delta}$) to those in Gaia DR2 in a 700 square degree region around ($\alpha$,$\delta$)=(45\degr,$-$30\degr).  The two datasets agree very well with the median offset in $\mu_\alpha$/$\mu_\delta$ being $-$0.248/$-$0.065 mas yr$^{-1}$ with a scatter of 2.45/2.36 mas yr$^{-1}$.

\begin{center}
\begin{figure*}[ht]
\includegraphics[width=0.47\hsize,angle=0]{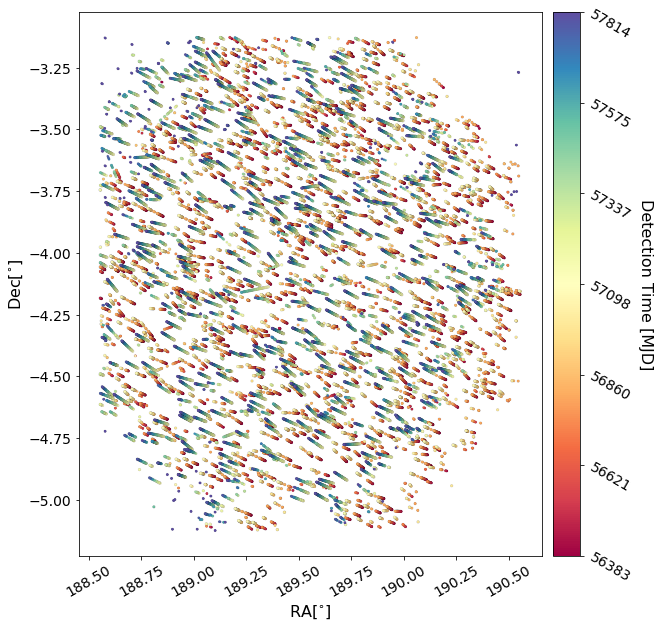}
\includegraphics[trim={-2cm -0.5cm 0cm 0cm},clip,width=0.5\hsize,angle=0]{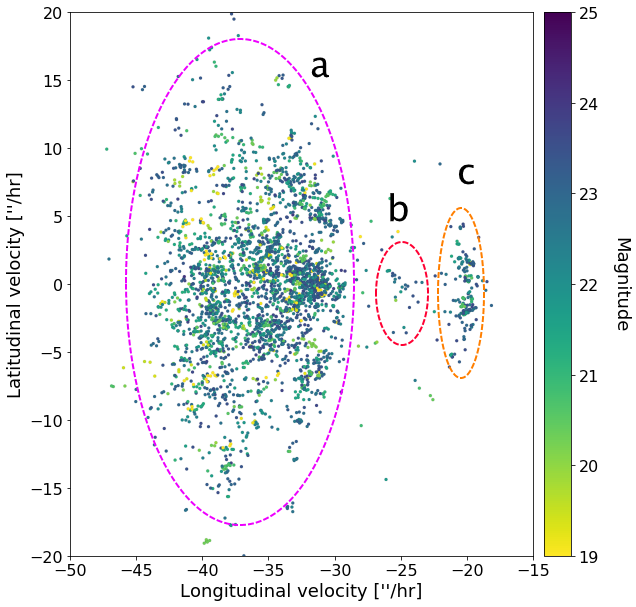}
\caption{\textit{(left)} ``Tracklets'' of solar system objects detected from the NSC in an area near the ecliptic plane, in equatorial coordinates [$^{\circ}$].  Individual measurements are color-coded by their observation time.  \textit{(right)} Proper motion [\arcsec/hr] of tracklets from left panel, in ecliptic coordinates.  3 groups of objects are shown: (a) Main Belt Objects, (b) Hilda asteroids, and (c) Jupiter Trojans.  }
\label{fig_tracklets}
\end{figure*}
\end{center}

NSC DR2 is being released through the NOIRLab's Astro Data Lab\footnote{\url{https://datalab.noirlab.edu}} \citep{Fitzpatrick2016,Nikutta2020}.  The database tables can be accessed via direct SQL queries using the Data Lab client software (Python) or via a TAP service\footnote{\url{http://datalab.noao.edu/tap}}.  The column descriptions can be viewed using the Data Lab query interface page\footnote{\url{https://datalab.noirlab.edu/query.php}}.  Data analysis and exploration can be performed using the Astro Data Lab's Jupyter Hub Notebook server running next to the data which provides fast access.

\section{Example Science Use Cases}
\label{sec:science}

There are many science use cases for a large catalog like the NSC DR2.  Below we describe a handful of them: Solar System objects (\S \ref{subsec:solarsystem}), stellar streams (\S \ref{subsec:stellarstreams}), variable stars (\S \ref{subsec:variables}), proper motion searches (\S \ref{subsec:propermotion}), and, QSO variability (\S \ref{subsec:qsos}).

\subsection{Solar System Objects}
\label{subsec:solarsystem}

The large temporal baseline and multiple repeat observations available in the NSC make it ideal for exploring Solar System objects (SSOs).  Figure \ref{fig_tracklets} shows 3,313 tracklets detection in the area of one DECam field near the ecliptic plane.  It is immediately obvious that a large fraction of the tracklets are in the same direction, reflecting the tendency of SSOs to have predominantly prograde orbits. 

Of all SSOs, identifying Near Earth Objects (NEOs) is of particular interest because of the danger they pose to the Earth.  Catastrophic effects can result from both large and small bodies; an asteroid with a diameter of 15 km likely caused the mass extinction event 65 million years ago that is widely believed to have killed a significant fraction of the non-avian dinosaurs, whereas the object that flattened 2,000 km$^2$ of forest in Tunguska in 1908 was ``only'' 200 m in diameter.  Concern over past and future impacts led the U.S. Congress to introduce the Spaceguard directive in the 1990's, directing NASA to find 90\% of NEOs with a diameters $\geq$1km \citep{Morrison}.  In 2011 NEOWISE \citep{Mainzer} reported the completion of the Spaceguard goal, and are now working towards the new goal of detecting 90\% of NEOs greater than 140 m in diameter\footnote{https://www.nasa.gov/planetarydefense/neoo} along with the Catalina Sky Survey (CSS), ATLAS \citep{atlas2018}, and LINEAR.  

The NSC expands the search carried out by projects such as the Palomar Transient Factory \citep{Law2009}, ZTF, CSS, and PS1, and the ones that will soon be possible with the Rubin Observatory's LSST.  In addition to being deeper than many existing surveys (and therefore able to detect smaller NEOs), the NSC adds data coverage in sparsely observed regions of the sky (e.g., the southern sky that PS1 does not reach and the Galactic plane that is unobserved by CSS). 

The NSC's depth suggests that it probably contains many detections of objects further from the sun.  Studying properties of the distant Kuiper belt objects (KBOs) will provide stronger constraints on planet formation theories, as KBOs are likely remnants of the primordial solar system.  Further detections of both KBOs and the even more distant Inner Oort Cloud objects can reveal the effects of external forces such as the Galactic tide, passing stars, or distant unknown planets \citep[such as the proposed Planet 9;][]{Sheppard2014} on our solar system and its formation history.  The Planet 9 hypothesis stems from an observed clustering in the orientation and phase of the orbits of distant solar system objects \citep{Sheppard2014}.  Although to date it remains undetected, the NSC data could contain detections of the elusive Planet 9---or additional distant solar system objects that could shed further light on Planet 9's existence.

Objects found in the NSC may also clarify the size distribution of solar system bodies.  The number of asteroids detected in groups of similar radii does not match the predictions \citep{Shep2010}, but the results of a search through the NSC data could alter the situation.
By investigating the range of asteroid sizes, new parameters will be established regarding the accretion and formation history of the solar system.  Therefore, with the right analysis techniques in hand, the deep, multi-band, time-series NSC information of 3.9 billion objects will allow us to markedly improve the census of solar system bodies and help our understanding of planet formation.

\begin{center}
\begin{figure}[t]
\includegraphics[width=1.0\hsize,angle=0]{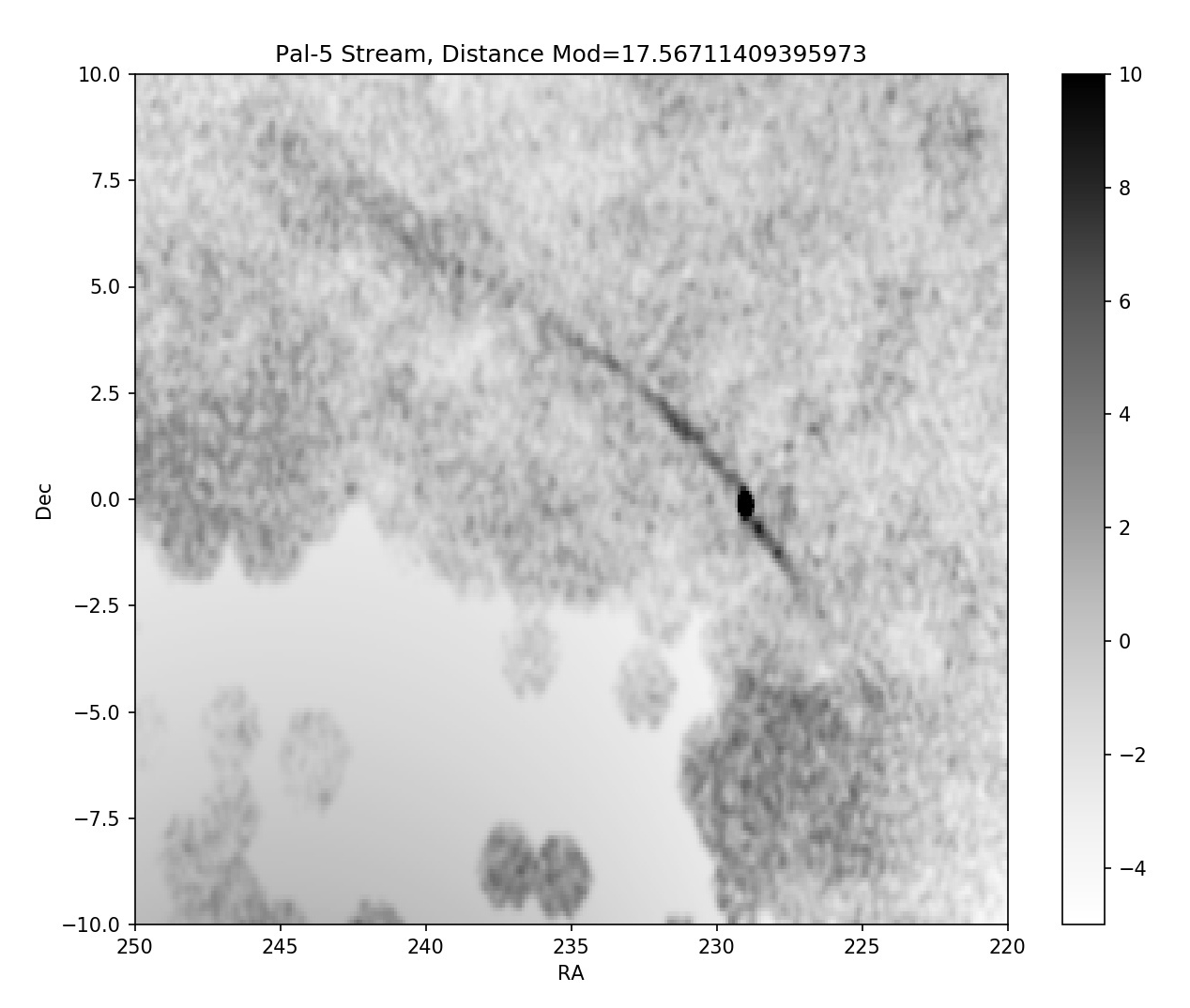}
\caption{The Palomar 5 stellar stream as seen in NSC DR2, which includes data used by \citet{Bonaca2020}.}
\label{fig_stream}
\end{figure}
\end{center}

\subsection{Stellar Streams}
\label{subsec:stellarstreams}

Stellar streams are the remnants of old globular clusters or dwarf galaxies that have been tidally disrupted and stretched apart by interactions with the Milky Way \citep[e.g, the Sagittarius stream;][]{Majewski2003,Koposov2012}. These linear over-densities of stars are very valuable for constraining the Galactic gravitational potential and cn potentially reveal dark matter sub-halos that disturb the otherwise uniform stream shape. With the knowledge that streams form from old star clusters and dwarf galaxies (which typically have a small range in stellar age), search algorithms can be tuned for these characteristics.
Using isochrones, we can search the sky for populations that fall within a small tolerance of these curves in color-magnitude space \citep[i.e., masked filters][]{Grillmair2006b}.  Stellar density maps at a large range of distance moduli can then be created and searched for linear overdensities, the tell-tale sign of a stream.
The broad spatial coverage and depth in multiple bands make the NSC DR2 very useful for detecting new stellar steams, especially in the southern hemisphere which has not yet been systematically searched the way the northern hemisphere has with SDSS and PS1.

Figure \ref{fig_stream} shows an example of the application of this technique to a region of sky near the well-known Palomar 5 stellar stream \citep[e.g.,][]{Odenkirchen2001,Grillmair2006a, Bonaca2020}. Searching the NSC DR2 with an isochrone with metallicity [Fe/H]=$-$0.5, an age of 11 Gyr, and distance modulus of 17.57 mag (33 kpc), the resulting density map clearly reveals the stream-like tidal tails of Pal 5. The NSC DR2 catalog covers new areas that haven't been searched extensively before and could reveal new stream candidates.

\subsection{Variable Stars}
\label{subsec:variables}

\begin{center}
\begin{figure}[t]
\includegraphics[width=1.0\hsize,angle=0]{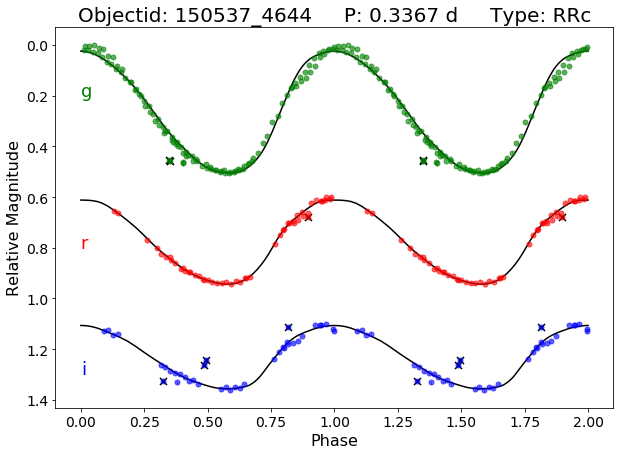}
\caption{Example lightcurve of an RR Lyrae star showing three bands.  Rejected outlier points are marked with $\times$s.}
\label{fig_rrlyrae}
\end{figure}
\end{center}

The NSC DR2's temporal baseline and depth are also very useful for detecting and studying variable stars, especially since the DR2 reports photometric variability metrics and an automatic selection of over 23 million variable objects.  Figure \ref{fig_rrlyrae} 
shows an example RR Lyrae lightcurve using data from NSC DR2.  Since variable stars are ``standard candles'', we can determine their distances accurately and use them as probes to study the structure of our Milky Way galaxy.
RR Lyrae variables, in particular, are plentiful and luminous and have been used for decades to explore the stellar structure of the Milky Way stellar halo.
\citet{Sesar2017b} used $\sim$40,000 RR Lyrae stars from PS1 to detect a new feature of the Outer Virgo Overdensity in the outer MW, while \citet{Hernitschek2017} used the same dataset to create an accurate 3D map of the Sgr stellar stream.
The deep NSC data can be used to detect RR Lyrae (and other variables) over nearly the entire sky and to larger distances than previously possible, extending these types of studies throughout the Milky Way and its satellites.

\subsection{Proper Motion Searches}
\label{subsec:propermotion}
\begin{center}
\begin{figure}[t]
\includegraphics[width=1.0\hsize,angle=0]{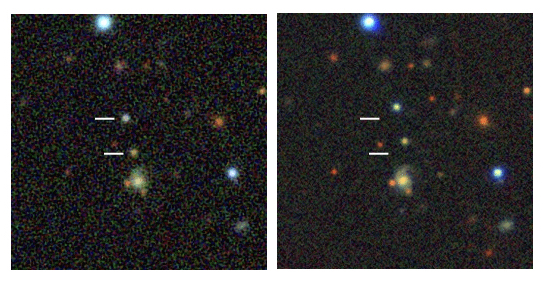}
\includegraphics[width=1.0\hsize,angle=0]{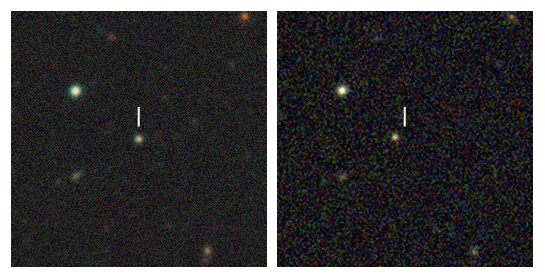}
\caption{Examples of high proper motion stars in NSC DR2. (Top) A co-moving pair of objects (21st and 22nd magnitude) with a proper motion of 270 mas/yr. (Bottom) A 21st magnitude star with a total proper motion of 205 mas/yr.  The Legacy Survey Viewer (\url{https://www.legacysurvey.org/viewer}) was used to generate the background RGB images.}
\label{fig_hpm}
\end{figure}
\end{center}

\textit{Gaia} DR2 has revolutionized astrometry, but the NSC nevertheless provides a valuable complement by providing proper motion measurements that push much fainter at optical wavelengths. At $g$ band, NSC is $\sim$2.5 magnitudes deeper than \textit{Gaia}. NSC DR2 will thus enable proper motion searches for distant stars with high tangential velocities over a volume $\sim$25 times larger than \textit{Gaia}, extending the many prior \textit{Gaia}-based studies of hypervelocity and runaway stars \citep[e.g.,][]{shen_hypervelocity_wd, kenyon14, brown_hypervelocity}. NSC can also measure motions for white dwarfs much fainter than those accessible to \textit{Gaia}, expanding the census of white dwarfs in the solar neighborhood. Accurate NSC proper motion measurements for faint white dwarfs will also help purify selections of faint quasars, and provide more opportunities to uncover valuable ultra-cool white dwarf binaries where metallicity and radial velocity can be obtained from a main sequence companion \citep[e.g.,][]{ucwd_benchmarking}.
Figure \ref{fig_hpm} shows two examples of high proper motion stars well detected in the NSC DR2 data.


By virtue of its excellent red-optical sensitivity and sky coverage, NSC DR2 will also provide many exciting opportunities to search for very late type stars and brown dwarfs in the solar neighborhood. CatWISE 2020 \citep{catwise_catalog} currently represents the best available infrared proper motion catalog, but NSC DR2 will offer capabilities not possible with CatWISE. At its faint end, CatWISE motions are only significant above $\sim$150--200 mas/yr. On the other hand, NSC measures motions many times smaller than this at high significance. Reliably identifying late type objects with low proper motions and accurately measuring those small motions are critical steps toward pinpointing young planetary mass brown dwarfs, such as those in nearby moving groups \citep[e.g.,][]{schneider_l_dwarfs}. NSC $Y$ band is also typically deeper than WISE for brown dwarfs in the late M to early T regime, whereas \textit{Gaia} is shallower than WISE for all brown dwarf types.

The $\sim$1$''$ angular resolution of NSC can also enable motion searches that are not feasible with WISE (which has FWHM $\sim$ 6$''$ from 3-5$\mu$m). For instance, NSC can be queried for pairs of faint/red objects with similar proper motions, to find closely spaced (few arcsecond separation) brown dwarf visual binaries. Similarly, NSC can be used to find close late-type co-moving companions to white dwarfs, providing valuable benchmark systems for the typically difficult task of estimating brown dwarf ages. In both of these examples, CatWISE would merely show one blended moving source rather than the resolved pair provided by NSC.


\subsection{QSO Variability}
\label{subsec:qsos}

Variations in the brightness of QSOs can be due to changes in the accretion disks and/or in the obscuration as dense absorbers might occult the central point source along our line of sight. 
Depending on its physical origin, QSO variability can occur over a range of timescales, with month-to-year long variations of $>$$1$~mag and  shorter timescale (days-to-weeks) variability as large as $>$$0.1$~mag. 
Variability measurements of QSOs are used to (1) identify them; and (2) infer physical properties (e.g., black hole masses from reverberation mapping, changes in accretion rates and/or obscuration).

\begin{center}
\begin{figure}[t]
\includegraphics[width=1.0\hsize,angle=0]{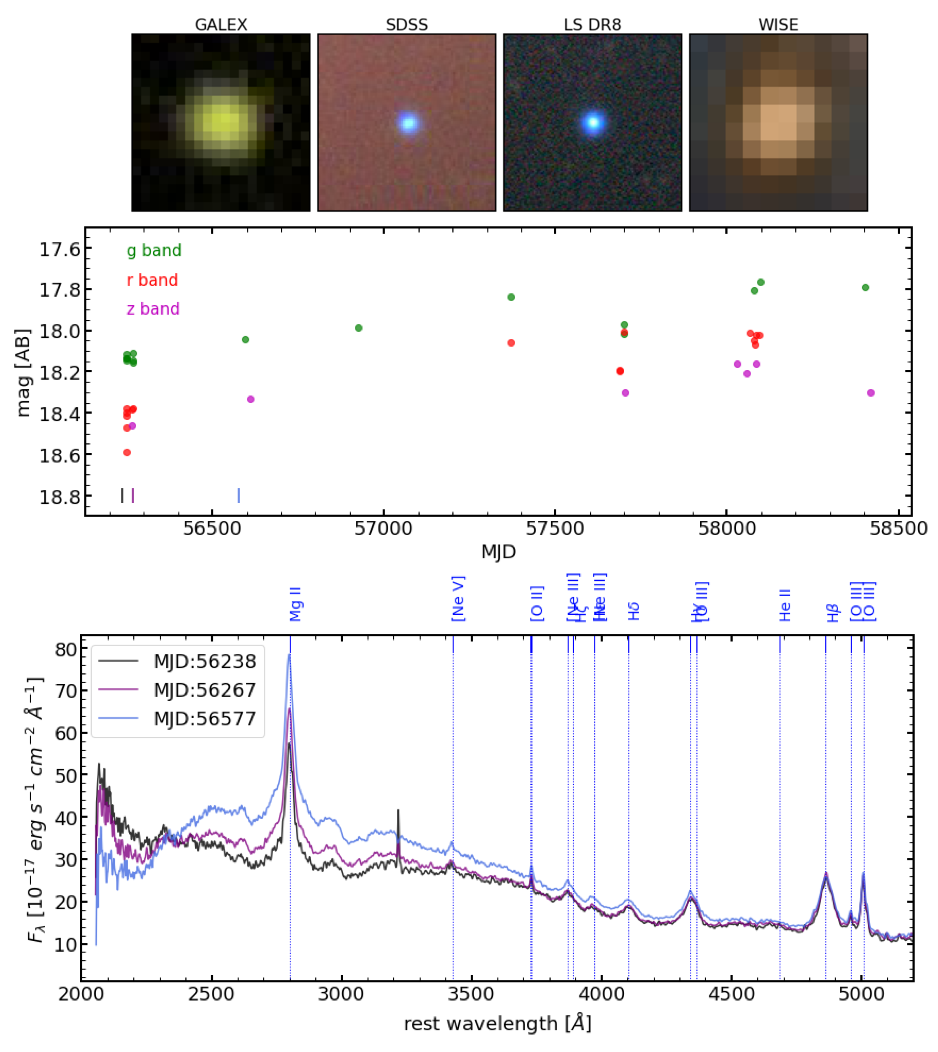}
\caption{(Top) Image cutouts (30 arcsec wide) of a $z\approx0.7$ variable QSO. (Middle) Lightcurve showing three bands as a function of the Modified Julian Date (MJD) of the observations. Vertical tick marks indicate when the spectra from the bottom panel were taken. (Bottom) SDSS spectra from three different MJDs, where the most striking differences are found in the spectral region around the Mg II line.}
\label{fig_qso1}
\end{figure}
\end{center}

Optical identification of QSOs typically relies on a point-source morphology (which may not strictly hold at low redshifts when the QSO host galaxy is resolved) and/or on color cuts to differentiate them from stars and galaxies. However, optical colors sometimes overlap between these various classes. Thus, using the unique signatures of QSO variability (which can be distinguished from stellar variability) can enable us to select samples of quasars across a range of redshifts. For instance, \citet{Palanque2011} found that QSO variability selection is more complete at $2.7<z<3.5$ compared to traditional optical color selections which suffer from overlap with stellar-like colors in this redshift range. Recently, researchers have used, e.g., SDSS Stripe 82 multi-epoch data \citep{Palanque2016}, Palomar Transient Factory \citep{Myers2015}, or the Catalina Real-time Transient Survey \citep{Graham2020} to search for QSOs based on variability. The NSC DR2 tends to reach fainter magnitudes than these datasets, but does not uniformly include as many epochs. Therefore, one could build from these previous efforts by comparing quasars that overlap, and devising a selection function tailored to the NSC measurements and pre-computed variability metrics.

\begin{center}
\begin{figure}[t]
\includegraphics[width=1.0\hsize,angle=0]{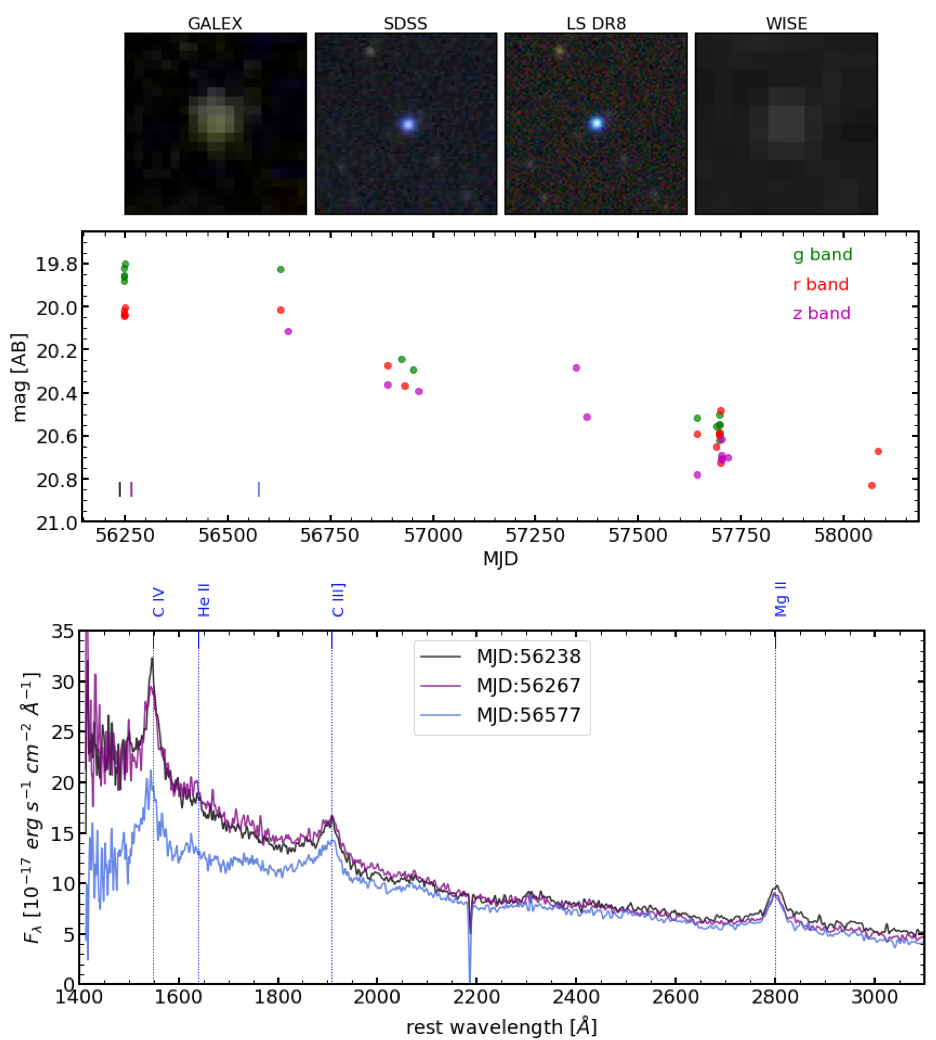}
\caption{(Top) As in Figure~\ref{fig_qso1}, but for a $z\approx1.55$ variable QSO. (Middle) Lightcurve showing three bands as a function of the MJD of the observations. Vertical tick marks indicate when the spectra from the bottom panel were taken. The overall trend indicates fading by $\sim$0.8~mag. (Bottom) SDSS spectra from three different MJDs. The last spectrum (light blue) displays fainter emission blueward of $\sim2000$~\AA.}
\label{fig_qso2}
\end{figure}
\end{center}

In addition to enabling population studies, the wide footprint of the NSC allows the search for rare sources like the {\it Changing-Look Quasars} (CLQ) or other sub-classes of AGN with the most extreme variations \citep[e.g.,][]{LaMassa2015,MacLeod2019}. As a proof of concept, we used the Astro Data Lab platform to cross-match the SDSS DR14Q quasar catalog \citep{Paris2018} with NSC DR2, finding a match within $1\arcsec$ for 527,552 quasars. We required NDET$>15$ to ensure a minimum sampling of NSC light-curves, yielding 133,013 quasars. We then examined cases with the most extreme variations (NSIGVAR$>10$), which also have multiple spectra from SDSS ($\geq$3~spectra). We show two different examples with obvious variations in both their SDSS spectra, and their NSC light curves in  Figures~\ref{fig_qso1} and \ref{fig_qso2}. The NSC is ideally suited to reveal many more interesting cases, and can further be extended beyond the SDSS footprint. It is sensitive enough to include QSOs out to higher redshifts ($z>2$), which is relevant given recent findings reporting the first cases of CLQs at $z>2$ \citep{Ross2020}. NSC photometric lightcurves will also complement future spectroscopic surveys such as the upcoming Dark Energy Spectroscopic Instrument (DESI) survey. 

\section{Summary}
\label{sec:summary}

We present the second public data release of the NOIRLab Source Catalog (NSC DR2) based on over 412,000 public images from the NOIRLab Astro Data Archive from both the northern and southern hemispheres.  The catalog contains 68 billion individual measurements to depths of $\approx$23rd magnitude of 3.9 billion unique objects across 86\% of the sky and over baselines of $\approx$7 years.  Due to the wealth of temporal information --- half a billion objects have 30 measurements or more --- the NSC DR2 delivers reliable proper motions (many stars fainter than the giant limit) as well as multiple photometric variability metrics.  The catalog enables a number of exciting science topics including (1) a census of Solar System bodies to faint depths, (2) searches for stellar streams and dwarf satellite galaxies in areas not previously probed, (3) cataloging variety types of variable stars, and (4) using QSO variability to identify and/or study these objects.

\acknowledgments

This project used data obtained with the Dark Energy Camera (DECam) at the Blanco 4m telescope at Cerro Tololo Inter-American Observatory. DECam was constructed by the Dark Energy Survey (DES) collaborating institutions: Argonne National Lab, University of California Santa Cruz, University of Cambridge, Centro de Investigaciones Energeticas, Medioambientales y Tecnologicas-Madrid, University of Chicago, University College London, DES-Brazil consortium, University of Edinburgh, ETH-Zurich, University of Illinois at Urbana-Champaign, Institut de Ciencies de l'Espai, Institut de Fisica d'Altes Energies, Lawrence Berkeley National Lab, Ludwig-Maximilians Universit\"at, University of Michigan, National Optical Astronomy Observatory, University of Nottingham, Ohio State University, University of Pennsylvania, University of Portsmouth, SLAC National Lab, Stanford University, University of Sussex, and Texas A\&M University. Funding for DES, including DECam, has been provided by the U.S. Department of Energy, National Science Foundation, Ministry of Education and Science (Spain), Science and Technology Facilities Council (UK), Higher Education Funding Council (England), National Center for Supercomputing Applications, Kavli Institute for Cosmological Physics, Financiadora de Estudos e Projetos, Funda\c{c}\~ao Carlos Chagas Filho de Amparo a Pesquisa, Conselho Nacional de Desenvolvimento Cientfico e Tecnol\'ogico and the Minist\'erio da Ci\^encia e Tecnologia (Brazil), the German Research Foundation-sponsored cluster of excellence "Origin and Structure of the Universe" and the DES collaborating institutions. The Cerro Tololo Inter-American Observatory, National Optical Astronomy Observatory is operated by the Association of Universities for Research in Astronomy (AURA) under a cooperative agreement with the National Science Foundation. 

This project also incorporates observations obtained at Kitt Peak National Observatory, National Optical Astronomy Observatory, which is operated by the Association of Universities for Research in Astronomy (AURA) under cooperative agreement with the National Science Foundation. The Kitt Peak data are largely drawn from the Mayall $z$-band Legacy Survey (MzLS), which was part of the Legacy Surveys project which imaged the footprint of the planned DESI survey. The Legacy Surveys imaging (which also included data taken using DECam) is supported by the Director, Office of Science, Office of High Energy Physics of the U.S. Department of Energy under Contract No. DE-AC02-05CH1123, by the National Energy Research Scientific Computing Center, a DOE Office of Science User Facility under the same contract. The paper also contains data from the Steward Observatory Bok 90" telescope, which is located on Kitt Peak and operated by the University of Arizona. The data obtained using the Bok telescope were obtained by the Beijing-Arizona Sky Survey, a key project of the Telescope Access Program (TAP), which has been funded by the National Astronomical Observatories of China, the Chinese Academy of Sciences (the Strategic Priority Research Program "The Emergence of Cosmological Structures" Grant \# XDB09000000), and the Special Fund for Astronomy from the Ministry of Finance. The BASS is also supported by the External Cooperation Program of Chinese Academy of Sciences (Grant \# 114A11KYSB20160057), and Chinese National Natural Science Foundation (Grant \# 11433005). The authors are honored to be permitted to conduct astronomical research on Iolkam Du'ag (Kitt Peak), a mountain with particular significance to the Tohono O'odham. 

This research uses services or data provided by the Astro Data Lab at NSF's National Optical-Infrared Astronomy Research Laboratory. NSF's NOIR Lab is operated by the Association of Universities for Research in Astronomy (AURA), Inc. under a cooperative agreement with the National Science Foundation.

This publication makes use of data from the Pan-STARRS1 Surveys (PS1) and the PS1 public science archive, which have been made possible through contributions by the Institute for Astronomy, the University of Hawaii, the Pan-STARRS Project Office, the Max-Planck Society and its participating institutes, the Max Planck Institute for Astronomy, Heidelberg and the Max Planck Institute for Extraterrestrial Physics, Garching, The Johns Hopkins University, Durham University, the University of Edinburgh, the Queen's University Belfast, the Harvard-Smithsonian Center for Astrophysics, the Las Cumbres Observatory Global Telescope Network Incorporated, the National Central University of Taiwan, the Space Telescope Science Institute, the National Aeronautics and Space Administration under Grant No. NNX08AR22G issued through the Planetary Science Division of the NASA Science Mission Directorate, the National Science Foundation Grant No. AST-1238877, the University of Maryland, Eotvos Lorand University (ELTE), the Los Alamos National Laboratory, and the Gordon and Betty Moore Foundation.

This work has made use of data from the European Space Agency (ESA)
mission {\it Gaia} (\url{https://www.cosmos.esa.int/gaia}), processed by
the {\it Gaia} Data Processing and Analysis Consortium (DPAC,
\url{https://www.cosmos.esa.int/web/gaia/dpac/consortium}). Funding
for the DPAC has been provided by national institutions, in particular
the institutions participating in the {\it Gaia} Multilateral Agreement.

This publication makes use of data products from the Two Micron All Sky Survey, which is a joint project of the University of Massachusetts and the Infrared Processing and Analysis Center/California Institute of Technology, funded by the National Aeronautics and Space Administration and the National Science Foundation.

Some of the results in this paper have been derived using the healpy and HEALPix package.


\software{
    \package{Astropy} \citep{astropy},
    \package{IPython} \citep{ipython},
    \package{matplotlib} \citep{mpl},
    \package{numpy} \citep{numpy},
    \package{scipy} \citep{scipy},
    \package{healpy} \citep{Zonca2019},
    \package{SExtractor} \citep{Bertin1996},
    \package{scikit-learn} \citep{scikit-learn}
}

\facilities{CTIO:Blanco (DECam), KPNO:Mayall (Mosaic-3), Steward:Bok (90Prime), Gaia, PS1, CTIO:2MASS, FLWO:2MASS, Sloan, Skymapper, WISE, Spitzer, GALEX, Astro Data Lab}


\bibliographystyle{aasjournals}
\bibliography{main.bib}



\end{document}

%% file: shorthand.tex
\newcommand{\dgr}{$^{\circ}~$}

%% file: authors.tex
\correspondingauthor{David L. Nidever}
\email{dnidever@montana.edu}
\author[0000-0002-1793-3689]{David L. Nidever}
\affiliation{Department of Physics, Montana State University, P.O. Box 173840, Bozeman, MT 59717-3840}
\affiliation{NSF's National Optical-Infrared Astronomy Research Laboratory, 950 North Cherry Ave, Tucson, AZ 85719}


\author[0000-0002-4928-4003]{Arjun Dey}
\affiliation{NSF's National Optical-Infrared Astronomy Research Laboratory, 950 North Cherry Ave, Tucson, AZ 85719}

\author{Katie Fasbender}
\affiliation{Department of Physics, Montana State University, P.O. Box 173840, Bozeman, MT 59717-3840}

\author[0000-0002-0000-2394]{St\'ephanie Juneau}
\affiliation{NSF's National Optical-Infrared Astronomy Research Laboratory, 950 North Cherry Ave, Tucson, AZ 85719}

\author{Aaron M. Meisner}
\affiliation{NSF's National Optical-Infrared Astronomy Research Laboratory, 950 North Cherry Ave, Tucson, AZ 85719}

\author{Joseph Wishart}
\affiliation{Department of Physics, Montana State University, P.O. Box 173840, Bozeman, MT 59717-3840}

\author{Adam Scott}
\affiliation{NSF's National Optical-Infrared Astronomy Research Laboratory, 950 North Cherry Ave, Tucson, AZ 85719}

\author{Kyle Matt}
\affiliation{Department of Physics, Montana State University, P.O. Box 173840, Bozeman, MT 59717-3840}

\author[0000-0002-7052-6900]{Robert Nikutta}
\affiliation{NSF's National Optical-Infrared Astronomy Research Laboratory, 950 North Cherry Ave, Tucson, AZ 85719}


\author{Ragadeepika Pucha}
\affiliation{Steward Observatory, University of Arizona, 933 North Cherry Ave, Tucson, AZ 85721}